\documentclass[11pt,a4paper]{book}
\usepackage{epsf,amsfonts,amsthm}
\usepackage{fontenc,indentfirst, delarray,amsfonts,amsmath,amssymb}
\usepackage{graphicx}
\DeclareGraphicsExtensions{ps,eps,ps.gz}

\epsfverbosetrue  
\makeatletter
\def\@citex[#1]#2{\if@filesw\immediate\write\@auxout
        {\string\citation{#2}}\fi
\def\@citea{}\@cite{\@for\@citeb:=#2\do
        {\@citea\def\@citea{,}\@ifundefined
        {b@\@citeb}{{\bf ?}\@warning
        {Citation `\@citeb' on page \thepage \space undefined}}
        {\csname b@\@citeb\endcsname}}}{#1}}
\newif\if@cghi
\def\cite{\@cghitrue\@ifnextchar [{\@tempswatrue
        \@citex}{\@tempswafalse\@citex[]}}
\def\citelow{\@cghifalse\@ifnextchar [{\@tempswatrue
        \@citex}{\@tempswafalse\@citex[]}}
\def\@cite#1#2{{\if@cghi\unskip$\null^{#1}$\else #1\fi\if@tempswa\typeout
        {warning: optional citation argument ignored: `#2'} \fi}}

\def\@biblabel#1{$\null^{#1}$}
\makeatother

\begin{document}
\newtheorem{lem}{Lemme}[chapter]
\newtheorem{thm}[lem]{Th\'eor\`eme}
\newtheorem{prop}[lem]{Proposition}
\newtheorem{cor}[lem]{Corollaire}
\newtheorem{defi}[lem]{D\'efinition}
\newtheorem{cond}{Condition}

\newcommand{\norm}[1]{\left\lVert#1\right\rVert}
\newcommand{\abs}[1]{\lvert#1\rvert}
\newcommand{\scal}[1]{\langle#1,#1\rangle}
\newcommand{\scl}[2]{\langle#1,#2\rangle}
\newcommand{\suup}[1]{ \underset{#1}{\sup} }
\newcommand{\grad}[1]{\text{grad}\,#1}
\newcommand{\ket}[1]{\lvert#1\rangle}
\newcommand{\bra}[1]{\langle#1\lvert}
\newcommand{\tr}[1]{\text{Tr}(#1)}

\def\re{\text{Re}}
\def\im{\text{Im}}

\def\dm{\lp\begin{array}}
\def\fm{\end{array}\rp}
\def\dbb{\lb\begin{array}}
\def\fbb{\end{array}\rb}
\def\dbn{\left.\begin{array}}
\def\fbn{\end{array}\right.}

\def\beq{\begin{equation}}
\def\eeq{\end{equation}}

\def\ee{{\cal E}}
\def\lb{\left[}
\def\rb{\right]}
\def\lp{\left(}
\def\rp{\right)}
\def\alg{alg\`ebre }
\def\nc{non commutative}
\def\gnc{g\' eom\' etrie \nc}
\def\ms{mod\`ele standard}

\def\calg{$C^*$-\alg}
\def\dn{\partial_{\nu}}
\def\ds{\slash\!\!\!\partial}
\def\da{\left[ D,\pi(a) \right]}
\def\df{\left[ D,f \right]}

\def\m3{M_3 \lp \cc \rp}
\def\m2{M_2 \lp \cc \rp}
\def\mmp{M_p \lp \cc \rp}
\def\mn{M_n \lp \cc \rp}
\def\mnp{M_{np} \lp \cc \rp}
\def\mpn{M_{pn} \lp \cc \rp}
\def\mx{M_k}
\def\mz{M_{k'}}
\def\kk{{\mathbb{K}}}
\def\cc{{\mathbb{C}}}
\def\rr{{\mathbb{R}}}
\def\nn{{\mathbb{N}}}
\def\zz{{\mathbb{Z}}}
\def\ii{{\mathbb{I}}}

\def\aa{{\cal A}}
\def\bb{{\cal B}}
\def\ll{{\cal L}}
\def\bbb{{\aa_I}}
\def\dd{{\cal D}}
\def\hh{{\cal H}}
\def\jj{{\cal J}}
\def\oo{{\cal O}}
\def\ss{{\cal S}}
\def\mm{{M}}
\def\pp{{\cal P}}
\def\ccc{{\cal C}}
\def\hhh{{\mathbb H}}
\def\jj{{\cal J}}
\def\gg{{\cal G}}
\def\xx{{\cal X}}
\def\ginf{\Gamma^\infty}

\def\cinf{C^{\infty}\lp\mm\rp}
\def \difm{\text{Diff}({\mathcal{M}})}
\def\cm{C^{\infty}(\mm)}
\def\spin{\text{Spin}}
\def\del{\triangledown}
\def\L2{L_2(\mm)}
\def\LS{L_2(\mm , S)}

\def\cnp{\cc^{np}}
\def\cpn{\cc^{pn}}
\def\cn{\cc^{n}}
\def\cp{\cc^{p}}

\def\ot{\otimes}
\def\pof{\psi\otimes\xi}
\def\fop{\xi\otimes\psi}

\def\xox{{\xi}_x}
\def\yox{{\xi}_y}
\def\xoz{{\zeta}_x}
\def\yoz{{\zeta}_y}

\def\omx{\omega_x}
\def\omy{\omega_y}

\def\ox{\omega_{\xi}}
\def\oz{\omega_{\zeta}}
\def\xo0{\omega^0_x}
\def\yo0{\omega^0_y}
\def\tox{\tilde{\omega_{\xi}}}
\def\toz{\tilde{\omega_{\zeta}}}
\def\oe{\omega_{E}}
\def\oi{\omega_{I}}
\def\oei{\omega_{E}\ot\omega_{I}}
\def\oeip{\omega_{E}\ot\omega_{I}'}
\def\oepi{\omega_{E}'\ot\omega_I }
\def\oepip{ {\omega_E}' \ot {\omega_I}'}
\def\xoi{\omega_x\ot\omega_i}
\def\xoip{\omega_x\ot\omega_{i}'}
\def\yoi{\omega_y\ot\omega_i}
\def\yoip{\omega_y\ot\omega_{i}'}

\def\ou{{\omega_{1}}}
\def\od{{\omega_{2}}}
\def\odu{\omega_{21}}
\def\oud{{\omega_{12}}}
\def\opud{{{\omega'}_{12}}}
\def\opdu{{{\omega'}_{21}}}
\def\ou{{\omega_{1}}}
\def\ouu{{\omega_{11}}}
\def\odd{{\omega_{22}}}
\def\oc{{\omega_{c}}}
\def\oeu{\omega_E \ot \ou}
\def\oed{\omega_E \ot \od}
\def\oeo{\omega_E \ot \omega}

\def\oxe{\omega_{k_e}}
\def\oze{\omega_{{k'}_e}}
\def\oue{\omega_{1}\circ\alpha_e}
\def\ode{\omega_{2}\circ\alpha_e}
\def\o0{\omega_0}
\def\rx{\rho_\xi}
\def\rz{\rho_\zeta}
\def\sx{s_\xi}
\def\sz{s_\zeta}

\def\xox{x_{k}}
\def\yox{y_{k}}
\def\rk{\rho_k}
\def\rkp{\rho_{k'}}
\def\ux{{u_k}}
\def\uz{{u_{k'}}}
\def\xoz{x_{{k'}}}
\def\yoz{y_{{k'}}}
\def\xo0{x_\omega^0}
\def\yo0{y_\omega^0}
\def\tox{\tilde{\omega_{k}}}
\def\toz{\tilde{\omega_{{k'}}}}
\def\nx{{n_k}}
\def\nz{{n_{k'}}}

\def\ae{\alpha_e}
\def\aea{\alpha_e(\aa)}
\def\fu{\alpha_u}
\def\fue{\alpha_{u^*}}
\def\tu{\tau_1}
\def\td{\tau_2}
\def\futu{\fu(\tu)}
\def\futd{\fu(\td)}
\def\fuetu{\fue(\tu)}
\def\fuetd{\fue(\td)}
\def\tue{\tu\circ\ae}
\def\tde{\td\circ\ae}

\def\te{\tau_E}
\def\ti{\tau_I}
\def\tei{\tau_E\ot\tau_I}
\def\po{\pi_\omega}
\def\pto{\tilde{\pi}_\omega}

\def\auta{\text{Aut}(\aa)}
\def\ina{\text{In(\aa)}}
\def\outa{\text{Out}(\aa)}\def\auta{\text{Aut}(\aa)}
\def\inmn{\text{In}(\mn)}
\def\outmn{\text{Out}(\mn)}

\def\cl{\text{Cl}}
\def\ccl{\cc\text{l}}

\title{Une br\`eve introduction \`a la description du mod\`ele
standard des particules \'el\'ementaires par la g\'eom\'etrie non
commutative}

\author{Pierre Martinetti\\[.2cm]
{\em martinetti@cpt.univ-mrs.fr}\\[.2cm]
notes d'un cours donn\'e \`a \\l'universit\'e Mohammed 1, Oujda,
Maroc, d'octobre 2002 à mars 2003,\\n dans le cadre d'un s\'ejour
de postdoctorat financ\'e par\\ l'Agence Universitaire de la
Francophonie.}
\date{\small\today} \maketitle

\section{Introduction}

La g\' eom\' etrie de notre espace pose probl\`eme en physique car
il n'en existe pas une description unique. Dans l'esprit de la
relativit\' e g\' en\' erale,  l'espace et le temps forment un
objet quadridimensionel dont la courbure est donn\' ee par la
distribution de masse. Quand un objet massif se d\'eplace, la
courbure change; la g\'eom\'etrie est un objet dynamique. Au
contraire la m\'ecanique quantique, et plus g\'en\'eralement la
th\'eorie quantique des champs, suppose la donn\'ee a priori d'un
espace dans lequel \'evoluent des champs. Pour reprendre une image
de [\citelow{rovelli}], la th\'eorie des champs prend l'espace
pour sc\`ene,  alors qu'en relativit\'e la sc\`ene elle-m\^eme
participe \`a l'action. La contradiction est d'autant plus
flagrante que chacune de ces th\'eories est valide et v\'erifi\'ee
avec pr\'ecision dans son domaine d'application: la gravitation
pour la relativit\'e; les interactions \' electromagn\'etiques,
faibles et fortes pour la th\' eorie quantique des champs. Cette
double approche de la g\'eom\'etrie n'est pas forc\' ement
scandaleuse. Rien n'interdit \`a deux descriptions de cohabiter,
tant que la cohabitation est harmonieuse. Mais les ph\'enom\`enes
qui rel\`event \`a la fois de la m\' ecanique quantique et de la
gravitation, comme le tout d\'ebut de l'univers dans la th\'eorie
du big-bang, ou l'effondrement gravitationel d'une \'etoile
pass\'ee une certaine \'echelle, brisent cette harmonie.
L'hypoth\`ese r\'epandue au jour d'aujourd'hui est, qu'\`a tout
petite \'echelle, aucune des descriptions g\'eom\'etriques
classiques n'est valable. La structure g\'eom\'etrique intime de
l'espace-temps n'est pas connue. Et la m\'ecanique quantique
sugg\`ere que l'hypoth\`ese du continu n'est pas justifi\'ee. On
estime que cette structure intime devrait \^etre visible \`a des
\'echelles de l'ordre de $10^{-33} \text{ cm}$. C'est la longueur
de Planck $l_p=\sqrt{\frac{{\cal G}\hbar}{c^3}}$ obtenue par
combinaison des constantes fondamentales $\cal{G}$ (constante de
Newton), $c$ (vitesse de la lumi\`ere), $\hbar$ (constante de
Planck). La g\'eom\'etrie non commutative\cite{connes}, en
\'etendant les concepts g\'eom\'etriques usuels de mani\`ere
compatible \`a la fois avec la relativit\'e g\'en\'erale et avec
la m\'ecanique quantique, propose des outils math\'ematiques pour
appr\'ehender la g\'eom\'etrie \`a cette \'echelle.

Pour l'heure bien entendu, aucune th\'eorie ne d\'ecrit l'univers
\`a cet ordre de pr\'ecision. Parmi les candidats au titre de
th\'eorie de la gravitation quantique, aucun n'a jusqu'\`a
pr\'esent franchi avec succ\`es le cap de la v\' erification
exp\'erimentale. Une approche naturelle consiste \`a quantifier le
champ gravitationel comme les autres champs, mais la th\'eorie
obtenue est non renormalisable, c'est \`a dire sans int\' er\^et
physique.
 N\' eamoins cette optique, amener la relativit\' e \`a la th\'eorie des champs, reste
valable et a suscit\' e (et suscite)  des travaux consid\'erables
qui, dans les raffinements les plus r\'ecents,  aboutissent \`a la
th\' eorie des cordes et la supersym\'etrie. L'unification est
obtenue mais aux prix d'hypoth\`eses physiques fortes: l'espace
temps est \`a 11 dimensions et il existe deux fois plus de
particules que celles connues jusqu'\`a pr\' esent (\`a chaque
particule connue correspond un partenaire supersym\'etrique). Pour
l'instant, aucune de ces hypoth\`eses n'a \'et\'e v\'erifi\'ee.
Cette approche de l'unification consid\`ere comme secondaire la
nature proprement g\'eom\'etrique de la relativit\'e g\'en\'erale
et s'inscrit plut\^ot dans la d\'emarche d'un Weinberg
expliquant\cite{weinberg}: {\it "... Einstein and his successors
have regarded the effects of a gravitational field as producing a
change in the geometry of space and time. At one time it was even
hoped that the rest of physics could be brought into a geometric
formulation, but this hope has met with disappointment, and the
geometric interpretation of the theory of gravitation dwindled to
a mere analogy, which lingers in our language in terms like
"metric", "affine connection", and "curvature", but is not
otherwise very useful. The important thing is to be able to make
predictions about images on the astronomers' photographic plates,
frequencies of spectral lines, and so on, and it simply doesn't
matter wheter we ascribe these predictions to the physical effect
of gravitational fields on the motion of planets  and photons or
to a curvature of space and time"}. Pour d'autres au contraire le
caract\`ere dynamique de la g\'eom\'etrie constitue l'apport
essentiel de la relativit\'e g\'en\'erale et toute la question
est, pr\'ecis\'ement, d'adapter cette dynamique g\'eom\'etrique au
contexte quantique. En clair, il s'agit d'affranchir la th\'eorie
quantique des champs d'un espace donn\'e a priori. On parle de
th\'eorie des champs "background independant", telle la "loop
quantum gravity"\cite{carlo}. Malheureusement cette th\'eorie pour
l'instant ne propose pas de tests exp\'erimentaux (pour une
\'etude compar\'ee des deux approches, de leurs succ\`es et
insuffisances, voir [\citelow{smolin}]).

La foi en "l'unification par la g\'eom\'etrie" se heurte \`a notre
mauvaise compr\'ehension de la th\'eorie des champs. En effet,
autant la relativit\'e g\'en\'erale a une interpr\'etation
g\'eom\'etrique simple, autant ce que dit la m\'ecanique quantique
de la g\'eom\'etrie n\'ecessite des \'eclaircis\-sements. Comment
d\'efinir un point de l'espace en m\'ecanique quantique ? Ou plus
exactement comment donner une signification physique \`a la notion
de point ? Une mani\`ere simple consiste \`a appeler point
l'endroit occup\'e par une particule \`a un instant donn\' e. Mais
\`a supposer que l'on connaisse avec pr\'ecision un point, les
relations d'incertitude de Heisenberg indiquent que l'on ne peut
conna\^{\i}tre avec pr\'ecision la position de la particule \`a un
autre instant. Autrement dit, si une particule permet de d\'efinir
un point, elle ne permet pas d'en d\' efinir un autre. Bien sur,
on peut consid\'erer plusieurs particules au m\^eme instant dont
on connait les positions avec pr\'ecision, et on d\'efinit ainsi
plusieurs points. Mais pour savoir comment ces points s'arrangent
les uns par rapport aux autres, pour {\it faire la g\'eom\'etrie},
il faut pouvoir mesurer des distances. Pour ce faire, il faut
qu'un m\^eme objet, par exemple l'une des particules, occupe \`a
un instant donn\'e le point $a$, et \`a un autre instant le point
$b$. Connaissant sa vitesse, on mesure son temps de vol et l'on en
d\' eduit la distance. Mais plus on saura avec pr\'ecision que la
particule occupe le point $a$ \`a l'instant $t$, moins on pourra
\^etre sur qu'elle occupe le point $b$ \`a l'instant suivant. La
m\'ecanique quantique sugg\`ere de raisonner sur des valeurs
moyennes. Le point
 est alors d\'efini comme la valeur moyenne \`a un instant donn\'e de l'observable position appliqu\'ee sur l'\'etat repr\'esentant la
particule. On op\`ere ainsi un changement de point de vue
important: le point n'est plus d\'efini en tant qu'objet abstrait
de la g\'eom\'etrie (tel qu'on l'apprend \`a l'\'ecole: "un point
n'a pas d'\'epaisseur, une ligne est un ensemble infini de
points"), c'est un objet alg\'ebrique, la valeur moyenne d'un
op\'erateur sur un \'etat.

Or les math\'ematiciens savent traduire en langage alg\'ebrique
les propri\'et\' es g\'eom\'etriques d'un espace. Plus
pr\'ecis\'ement, les propri\'et\'es g\'eom\'etriques
(essentiellement la topologie, la mesure et la m\'etrique) d'un
espace ont une  traduction alg\'ebrique dans l'ensemble des
fonctions, \`a valeur complexe, d\'efinies sur cet espace. Par
exemple, la distance entre deux points $x,y$ est la longueur du
plus court chemin reliant $x$ \`a $y$. Mais c'est aussi le
supr\'emum, parmi toutes les fonctions dont la d\' eriv\'ee (en
valeur absolue) est toujours inf\' erieure \`a $1$,  du module de
la diff\'erence $f(x)-f(y)$. Ceci se v\'erifie sans difficult\'e
sur un exemple simple. Choisissons comme espace la droite
r\'eelle. La fonction $f$ d\' efinie sur $\rr$ par $f(x)=x$ a une
d\'eriv\' ee constante $f'(x)=1$, et on a bien
$$\abs{f(x)-f(y)}=\abs{x-y}= \text{distance}(x,y).$$
Si une fonction
$g$ est telle que $\abs{g(x)-g(y)}>\abs{x-y}$, alors par le th\'
eor\`eme de la valeur interm\' ediaire il existe n\' ecessairement
un r\'eel $c\in[x,y]$ tel que
$$\abs{g'(c)}=\frac{\abs{g(x)-g(y)}}{\abs{x-y}}>1.$$
On voit ainsi
que les deux d\' efinitions de la distance, l'une comme plus court
chemin, l'autre comme supremum d'une diff\'erence d'observables,
co\"{\i}ncident.

Cet exemple \'el\'ementaire illustre comment faire de la g\'
eom\'etrie de mani\`ere alg\'ebrique. Plus g\'en\'eralement la
g\'eom\'etrie au sens usuel est commutative, c'est \`a dire que
son expression alg\'ebrique prend pour cadre la th\'eorie des
alg\`ebres commutatives. Rappelons qu'une alg\`ebre est un
ensemble muni d'une loi d'addition et de multiplication  par un
scalaire, sur lequel est d\' efini en outre une multiplication.
Dans l'ensemble des fonctions \`a valeur complexe sur un espace,
ces lois sont d\'efinies point par point. Pour la multiplication
par exemple, si $f$ et $g$ sont deux fonctions sur un espace $X$,
alors
$$(f.g)(x)\doteq f(x).g(x)=g(x).f(x)=(g.f)(x).$$
Parce que le produit de deux nombres complexes est commutatif, le
produit de deux fonctions est commutatif, c'est \`a dire que
l'alg\`ebre des fonctions sur un espace est commutative.
Inversement,  \' etant donn\' ee une alg\`ebre commutative $\aa$,
on sait construire (construction de Gelfand-Naimark-Segal) un
espace $M$ tel que $\aa$ soit l'alg\`ebre des fonctions
(continues) sur $M$. Ainsi il est \' equivalent de se donner un
espace ou une alg\`ebre commutative: les propri\'et\' es g\'
eom\'etriques d'un espace ont une traduction dans l'alg\`ebre des
fonctions sur cet espace, et inversement les propri\'et\'es alg\'
ebriques d'une alg\`ebre commutative ont une traduction dans
l'espace associ\'e par la construction GNS:
$$
\text{espace } \Longleftrightarrow \text{alg\`ebre commutative.}
$$
La question naturelle est
$$
\text{?} \Longleftrightarrow \text{ alg\`ebre non commutative.}
$$

Naturellement, on ne saurait construire un espace tel qu'une
alg\`ebre non commutative soit son alg\`ebre de fonctions, puisque
l'alg\`ebre des fonctions sur un espace est n\' ecessairement
commutative. La g\'eom\'etrie non commutative est une adaptation
du dictionnaire qui permet de passer "d'alg\`ebre commutative" \`a
"espace" en remplacant, partout o\`u il y a lieu, le mot
commutatif par non commutatif. Evidemment les choses ne sont pas
si simples. Abandonner la commutativit\'e implique de profonds
changements dans les d\' efinitions du dictionnaire, et requiert
m\^eme la cr\' eation de notions nouvelles. L'investissement
math\'ematiques est lourd mais le jeu en vaut la chandelle car on
peut alors acc\'eder \`a de nouveaux types "d'espaces non
commutatifs" o\`u des ph\' enom\`enes physiques trouvent une
interpr\'etation g\'eom\'etrique qu'ils n'avaient pas jusque l\`a.
Par exemple le champ de Higgs apparait comme le coefficient d'une
m\'etrique dans une dimension suppl\'ementaire, discr\`ete, qui
rend compte des degr\'es de libert\'e internes (spin ou isospin)
d'une particule.
\newline

Dans le chapitre suivant, on rappelle comment un espace au sens
usuel est topologiquement \'equivalent \`a une alg\`ebre
commutative, ce qui permet d'identifier les \'etats purs d'une
alg\`ebre comme \'equivalent non commutatif \`a la notion
classique de "point". Le deuxi\`eme chapitre pr\'esente la
structure diff\'erentielle due \`a Connes, en particulier le
th\'eor\`eme fondamental qui - dans sa version commutative -
fournit une d\'efinition axiomatique d'une vari\'et\'e \`a spin,
et propose - dans sa version non commutative - la notion de {\it
triplet spectral r\'eel} comme outil d'investigation de l'univers
non commutatif. Le troisi\`eme chapitre s'int\'eresse \`a la
formule de la distance, en s'attardant sur la distance
g\'eod\'esique dans une vari\'et\'e \`a spin ainsi que sur divers
exemples d'espaces non commutatifs finis. Dans le quatri\`eme
chapitre, on aborde de mani\`ere succinte la description du
mod\`ele standard des particules \'el\'ementaires, en mettant
l'accent sur l'interpr\'etation du champ de Higgs comme
coefficient de la m\'etrique dans une dimension suppl\'ementaire
discr\`ete. Enfin le dernier chapitre \'etudie la question des
neutrinos massifs.
\newline

On emploie la convention d'Einstein de sommation sur des indices
r\'ep\'et\'es, uniquement en position altern\'ee (haut-bas).

\chapter{Topologie de l'espace non commutatif}

Avant de pr\'eciser ce qu'est un espace non commutatif, il n'est
pas inutile de rappeler en quoi un espace g\'eom\'etrique, au sens
usuel, est un espace commutatif. Ceci permet l'introduction des
{\it \'etats purs} comme "points" de l'espace non commutatif. On
trouvera les d\'emons\-trations dans des trait\'es d'alg\`ebres
d'op\'erateurs tels que [\citelow{takesaki,kadison,murphy}] ou
dans [\citelow{jgb}] pour un traitement plus orient\'e vers la
g\'eom\'etrie non commutative.

\section{Equivalence topologique entre espace usuel et alg\`ebre
commutative}

Au sens le plus \'el\'ementaire, faire de la g\'eom\'etrie c'est
\^etre capable de d\'eterminer si deux  \'el\'ements sont voisins
l'un de l'autre. C'est en effet sous cette condition qu'un
ensemble prend le nom d'espace. Math\'ematiquement, il s'agit de
munir un ensemble $X$ d'une topologie, c'est \`a dire de d\'efinir
la notion de sous-ensemble ouvert (d'o\`u celle de fonction
continue). Quand  la topologie est suffisamment fine pour
distinguer les points, $X$ est dit s\' epar\'e (ou Hausdorff).
 $X$ est compact signifie que de tout recouvrement infini d'ouverts $U_i$ \-- $\underset{i=1}{\overset{\infty}\bigcup} U_i = X$\-- on
peut extraire un recouvrement fini. On observe alors que
l'ensemble $C(X)$ \index{cx@$C(X)$} des fonctions \`a valeur
complexe continues sur $X$ est une alg\`ebre complexe commutative
qui, en tant qu'espace vectoriel, est compl\`ete pour la
m\'etrique induite par la norme
\begin{equation}
\label{normope} \norm{f}\doteq \suup{x\in X} \abs{f(x)}
\end{equation}
($C(X)$ est un espace de Banach).
 En tant qu'alg\`ebre $C(X)$ est munie d'une involution $^*$ naturelle h\'erit\'ee de la conjugaison complexe ainsi que d'une unit\'e (la fonction constante $1$). La
 norme v\'erifie
 $$\norm{fg}\leq\norm{f}\norm{g}$$
 ($C(X)$ une \alg de Banach) ainsi que
\begin{equation}
\label{c*} \norm{f}^2= \norm{f f^*}
\end{equation}
($C(X)$ est une $C^*$-alg\`ebre\index{cetoile@$C^*$}). A tout
espace topologique compact se trouve donc associ\'ee de mani\`ere
canonique une $C^*$-alg\`ebre complexe commutative avec unit\'e.

R\'eciproquement, \`a toute \calg complexe $\aa$ commutative
correspond l'espace localement  compact (pour la topologie
*faible) $K(\aa)$ \index{ka@$K(\aa)$} des caract\`eres de $\aa$.
Un caract\`ere est un homomorphisme d'\alg (n\'ecessairement
surjectif)  $$\mu: \aa \rightarrow \cc.$$ Soulignons plusieurs
propri\'et\'es
(\ref{caracun},\ref{caracspec},\ref{caracnorm},\ref{caracinvolution})
des caract\`eres, simples mais essentielles en ceci qu'elles
constituent le pivot de la g\'en\'eralisation au cas non
commutatif. Tout d'abord lorsque $\aa$ poss\`ede une unit\'e
$\ii$, $\mu(\ii)=\mu(\ii)^2$ d'o\`u
\begin{equation}
\label{caracun} \mu(\ii)= 1.
\end{equation}
Il s'en suit qu'un \'el\'ement inversible ne peut avoir pour image
z\'ero; donc $a - \mu(a)\ii$ n'est pas inversible. Autrement dit,
pour tout caract\`ere $\mu$ et tout $a$ de $\aa$,
\begin{equation}
\label{caracspec} \mu(a) \in \text{ sp} (a)
\end{equation}
ou $\text{ sp} (a)$, \index{spa@ $\text{sp} (a)$} le spectre de
$a$, est l'ensemble des valeurs $\lambda$ telles que
$a-\lambda\ii$ n'est pas inversible. Ensuite, sachant que pour
tout \'el\'ement $a$ d'une $C^*$-\alg complexe
\begin{equation}
\label{normspec} \suup{\lambda\, \in \text{ sp} (a)}
\abs{\lambda}\leq \norm{a},
\end{equation}
l'\'egalit\'e \'etant atteinte pour les \'el\'ements normaux
($a^*a = aa^*$), on observe que
\begin{equation}
\label{caracnorm} \norm{\mu}\doteq \suup{a\in\aa}
\frac{\abs{\mu(a)}}{\norm{a}}=1.
\end{equation}
Enfin, on montre qu'un caract\`ere \'evalu\'e sur un \'el\'ement
autoadjoint a valeur dans $\rr$. En d\'ecomposant tout $a$ en
\'el\'ements autoadjoints, $a = a_1 + i a_2$ avec $a_1 \doteq
\frac{1}{2}(a^* + a)$ et $a_2\doteq \frac{i}{2}(a^* - a)$, il
apparait qu'un caract\`ere pr\'eserve l'involution
\begin{equation}
\label{caracinvolution} \mu(a^*) = \mu(a_1 - i a_2) = \mu(a_1) -
i\mu(a_2) =  \bar{\mu}(a).
\end{equation}
Une forme lin\'eaire de ce type est dite {\it involutive}.

A l'aide de ces propri\'et\'es, on \'etablit (th\'eor\`eme de
Gelfand) que la transformation qui \`a tout $a\in\aa$ associe
l'application $\hat{a}\in C_0(K(\aa))$\index{achapeau@$\hat{a}$},
$$\hat{a}(\mu)\doteq \mu(a),$$
est un *isomorphisme isom\'etrique (i.e. pr\'eservant l'involution
et la norme) de $\aa$ dans  $C_0(K(\aa)).$ Lorsque $\aa$ est munie
d'une unit\'e, $K(\aa)$ est compact et $\aa$ est *isomorphe \`a
l'ensemble des fonctions continues sur $K(\aa)$. Quand une \alg
n'a pas d'unit\'e, on peut toujours lui en adjoindre une en
consid\'erant l'alg\`ebre augment\'ee. {\bf On suppose donc
dor\'enavant, sauf mention contraire, que les alg\`ebres ont une
unit\'e $\ii$.}\index{i@$\ii$} Avec cette convention,  le
th\'eor\`eme de Gelfand signifie que toute $C^*$-alg\`ebre
complexe commutative peut-\^etre vue comme l'alg\`ebre des
fonctions continues sur son espace des caract\`eres.
\newline

Ainsi \`a toute $C^*$-\alg complexe commutative $\aa$ est
associ\'e un espace compact $K(\aa)$, tandis qu'\`a tout espace
compact $X$ est associ\'e une *\alg commutative $C(X)$.
 Le th\'eor\`eme de Gelfand assure que
$$\aa \longrightarrow K(\aa) \longrightarrow C(K(\aa))\sim \aa.$$
A l'inverse on montre que l'espace des caract\`eres de $C(X)$
n'est autre que $X$,
$$X \longrightarrow C(X)  \longrightarrow K(C(X))\sim X.$$
Dans un langage plus rigoureux\cite{jgb}, la cat\'egorie des
$C^*$-alg\`ebres commutatives complexes avec unit\'e est
\'equivalente \`a la cat\'egorie (oppos\'ee) des espaces
compacts,
$$
\text{espace compact} \Longleftrightarrow \text{$C^*$-alg\`ebre
commutative.}.
$$
Sans entrer dans le d\'etail du langage des cat\'egories,
soulignons l'importante cons\'equence de cette \'equivalence:

{\prop \label{homeoiso}
 Deux $C^*$-\alg complexes commutatives sont isomorphes si et seulement si leurs espaces de carac\-t\`eres sont hom\'eomorphes.}
\newline

\noindent De mani\`ere plus g\'en\'erale, toute l'information
topologique d'un espace compact est contenue dans $C(X)$.
Soulignons que ceci reste vraie pour l'alg\`ebre des fonctions
lisses $\cinf\index{cinf@$\cinf$}$ sur une vari\'et\'e compacte
$\mm$: bien que $\cinf$ ne soit pas une $C^*$-alg\`ebre mais
seulement une sous-alg\`ebre dense de $C(\mm)$, tout caract\`ere
de $\cinf$ s'identifie \`a un point de $\mm$.

En cons\'equence deux points de vue sont possibles: classiquement
on prend les points $x$ comme objet premier et on interpr\`ete les
r\'esultats de l'exp\'erience comme des \'evaluations
d'observables sur ces points, ou bien on consid\`ere les
observables $f$ comme premi\`eres et les points sont, par
d\'efinition, les objets \'evaluant les observables. Quand
l'espace peut \^etre munie d'une topologie, c'est \`a dire quand
les observables (vues comme fonctions continues) commutent, ces
deux points de vue sont \'equivalents,
$$x(f) = f(x),$$
et les points sont les caract\`eres de l'\alg des observables.
Mais en m\'ecanique quantique la partie droite de l'\'equation,
l'\'evaluation d'une observable en un point, est mal d\'efinie. En
revanche l'ensemble des observables est bien d\'efini et c'est une
\alg non commutative. Pour donner sens \`a la partie gauche de
l'\'equation, il suffit de trouver l'objet \'equivalent au
caract\`ere pour une \alg non commutative.

 \section{Etats purs}

Lorsque $\aa$ est une $C^*$-alg\`ebre complexe non commutative,
ses caract\`eres ne forment pas un ensemble localement compact.
Ils ne sont d'ailleurs pas int\'eressants en regard de la non
commutativit\'e puisqu'un caract\`ere, par nature, identifie $ab$
\`a $ba$ ($ab - ba$ a pour image zero). N\'eammoins, \`a la
lumi\`ere du th\'eor\`eme de Gelfand, les $C^*$-alg\`ebres non
commutatives sont le candidat id\'eal pour jouer le r\^ole d'\alg
des fonctions d'un "espace non commutatif". En tant qu'ensemble,
cet espace est compos\'e des formes lin\' eaires sur l'alg\`ebre
qui v\'erifient
 les propri\'et\'es des caract\`eres, except\'ees celles ayant trait \`a la commutativit\'e (\`a savoir la multiplicativit\'e:  $\mu(ab)=\mu(a)\mu(b)= \mu(b)\mu(a)=\mu(ba)).$

{\defi \label{etat} Un \' etat sur une $C^*$-alg\`ebre complexe
est une forme $\cc$-lin\' eaire positive de norme $1$.}
\newline

\noindent La norme est la norme d'op\'erateur d\'efini en
(\ref{normope}).

On rappelle qu'un \' el\' ement $a$ est {\it positif} s'il est
autoadjoint et $\text{sp(a)}\subset [0,+\infty[$ ou, de mani\`ere
\'equivalente, s'il existe un \' el\' ement $b$ tel que $a=b^*b$.
L'ensemble des \' el\'ements positifs est not\'e
$\aa_+$\index{aplus@$\aa_+$} et une forme lin\'eaire
$\tau$\index{tau@$\tau$} est positive si
$\tau(\aa^+)=\cc^+=\rr^+$. On montre [\citelow{kadison}, {\it Th.
4.3.2}] qu'une forme lin\'eaire $\tau$  sur une \alg de Banach
avec unit\'e est positive si, et seulement si,  elle est born\'ee
et $\norm{\tau}=\tau(\ii)$. Par cons\'equent un \'etat se
d\'efinit de mani\`ere \'equivalente comme une forme
$\cc$-lin\'eaire positive satisfaisant
\begin{equation}
\label{etatun} \tau(\ii) = 1,
\end{equation}
ou encore comme une forme $\cc$-lin\'eaire born\'ee telle que
\begin{equation}
\label{etatequiv} \norm{\tau} = \tau(\ii) = 1.
\end{equation}
La positivit\'e est une condition n\'ecessaire mais non suffisante
pour garantir l'involutivit\'e. Cependant quand l'alg\`ebre a une
unit\'e la positivit\'e implique [\citelow{takesaki}, {\it Lem.
9.11}], et donc \'equivaut \`a,
\begin{equation}
\label{etatinvolution} \tau(a^*)= \bar{\tau}(a).
\end{equation}


L'espace des \'etats est convexe. Les points extr\'emaux, c'est
\`a dire les \'etats $\tau$ pour lesquels il n'existe pas
d'\'etats $\tau_1$, $\tau_2\neq \tau$ et de nombre $t\in [0,1]$
tels que  $\tau= t\tau_1 + (1-t)\tau_2$, sont appel\' es {\it
\'etats purs}.  Dans le cas commutatif, les caract\`eres
s'identifient aux \'etats purs. Par analogie ce sont les \'etats
purs de $\aa$, not\'e $\pp(\aa)$\index{pa@$\pp(\aa)$}, qui
tiennent lieu de "points" pour l'espace non commutatif,
$$
\text{espace non commutatif} \Longleftrightarrow \text{
$C^*$-alg\`ebre non commutative.}
$$
Il s'agit d'une analogie, non d'une d\'efinition stricte. Pour
certains r\'esultats (en particulier concernant les distances), on
est amen\'e en prendre en compte des \'etats non purs.


 Les \'etats de $\aa$  constituent le socle de l'espace non commutatif parce qu'ils y jouent le
 r\^ole des "points", mais aussi parce qu'ils garantissent, par la construction GNS (Gelfand-Naimark-Segal),
de pouvoir travailler concr\`etement avec $\aa$ vue comme sous
alg\`ebre de l'alg\`ebre des op\'erateurs born\' es sur l'espace
de Hilbert
{\thm \label{gns} Toute $C^*$-\alg complexe a une repr\'esentation
isom\'etrique en tant que sous-alg\`ebre de l'\alg $\bb(\hh)$ des
op\'erateurs born\'es sur un espace de Hilbert.}
%
%
\newline

\noindent On renvoie aux trait\'es d'alg\`ebre pour une \'etude de
la construction GNS. Soulignons simplement que ce r\'esultat est
fondamental puisqu'il permet, en consid\'erant des
$C^*$-alg\`ebres, de travailler concr\`etement avec une alg\`ebre
d'op\'erateurs sur un espace de Hilbert.

\chapter{Structure diff\'erentielle pour l'espace non commutatif}

 Le th\'eor\`eme de Gelfand \'etablit une \'equivalence entre
alg\`ebre commutative et espace topologique. Par analogie on
"\'etend" cette \'equivalence aux alg\`ebres non commutatives en
interpr\'etant ces derni\`eres comme alg\`ebres des fonctions sur
un "espace non commutatif" dont les points sont donn\'es par les
\'etats purs. Mais une analogie topologique n'est pas suffisante.
Outre la topologie, l'espace physique est muni d'une structure
diff\'erenti\-elle et d'un espace de degr\'es de libert\'e
internes (le spin en m\'ecanique quantique, l'isospin pour le
mod\`ele standard). De m\^eme que les propri\'et\'es topologiques
peuvent \^etre traduites en terme d'alg\`ebre commutative, l'objet
math\'ematique utilis\'e pour d\'ecrire l'espace de la th\'eorie
quantique des champs \- la {\it vari\'et\'e \`a spin} \- a une
d\'efinition alg\'ebrique. Privil\'egier cette d\'efinition rend
possible son adaptation aux espaces non commutatifs. Apr\`es
quelques rappels de g\'eom\'etrie diff\'erentielles, on pr\'esente
ici la construction de la structure de spin (puis de l'op\'erateur
de Dirac) par les modules de Clifford (cf. [\citelow{jgb}] pour un
expos\'e d\'etaill\'e) plut\^ot que la construction en fibr\'e
principal souvent d\'evelopp\'ee\cite{choquet,jost,nakahara} car
la transition au domaine non commutatif est alors plus ais\'ee.
Cette transition, via les axiomes de la g\'eom\'etrie non
commutative\cite{gravity} le th\'eor\`eme de Connes sont donn\'es
en fin de section. Inutile de pr\'eciser qu'il ne s'agit l\`a que
d'un survol de la th\'eorie, on renvoie \`a [\citelow{connes}]
pour l'expos\'e fondamental, ainsi qu'\`a [\citelow{jgb}] pour une
pr\'esentation d\'etaill\'ee des preuves.

\section{Vari\'et\'es, fibr\'es, m\'etrique}

On d\'ecrit l'espace physique (ou l'espace temps en relativit\'e)
par une vari\'et\'e diff\'erentiable $M$ de dimension $m$, sur
laquelle on d\'efinit un syst\`eme de coordonn\'ees locales
$x^\mu$, $\mu=1,...,m$. Une courbe $c$ dans $M$ est la donn\'ee de
$m$ fonctions coordonn\'ees $c^\mu: t\mapsto c^{\mu}(t)\in \rr$.
Le {\it vecteur} $X(x_0)$, tangent \`a la courbe $c$ en
$x_0=c(0)$, est d\'efini par son action sur une fonction $f:
M\rightarrow \rr,$
$$X[f]\doteq \frac{d}{dt} f(c(t)) =
\frac{\partial}{\partial_\mu}f(c^{\mu}(t))\lvert_{t=0} \frac{d
c^{\mu}}{dt}\lvert_{t=0}.$$ En associant de la sorte un vecteur
\`a tout point $c(t)$, on d\'efinit le champ de vecteur tangent
$$X = X^{\mu}\frac{\partial}{\partial_\mu}
$$
o\`u $X^{\mu}(c(t)) = \frac{d c^{\mu}}{dt}\lvert_{t}.$ En tout
point $x$ de $M$, l'ensemble des vecteurs tangents aux courbes
passant par $x$ forme un espace vectoriel: l'espace tangent $T_xM$
dont la base canonique en coordonn\'ees locales est not\'ee
$\{\partial_\mu\doteq \frac{\partial}{\partial_\mu}\}$. L'espace
cotangent, dual de l'espace tangent, est l'espace des {\it formes
differentielles} dont la base canonique $\{dx^{\mu}\}$ est
donn\'ee par la relation de dualit\'e
$$\scl{\partial_\mu}{dx^\nu} = \delta_\mu^\nu$$
o\`u $\delta$ d\'esigne le symbole de Kronecker.

Soit $\kk$ un corps (typiquement $\rr$ o\`u $\cc$). Un
$\kk$-fibr\'e vectoriel $E \overset{\pi}\longrightarrow M$ est un
espace topologique localement hom\'eomorphe au produit $U_i\times
F$, o\`u $U_i$ est un ouvert de $\mm$ et $F$ un espace vectoriel
sur $\kk$, $\pi$ d\'esignant la projection de $E$ sur $M$. Pour
tout $x$ de $\mm$, $E_x \doteq \pi^{-1}(x)$\index{ex@$E_x$}, la
fibre au dessus de $x$, est isomorphe \`a $F$. Une section locale
$\sigma_i$\index{sigmai@$\sigma_i$} de $E$ est une application de
$U_i$ dans $E$ telle que $\pi\circ\sigma_i$ soit l'identit\'e de
$U_i$. Si $r$ est la dimension de $F$, une section locale est la
donn\'ee de $r$ fonctions de $U_i$ dans $\rr$, appel\'ees
composantes de la section.  Une section locale est
diff\'erentiable (continue) quand ces composantes sont des
fonctions diff\'erentiables (continues) sur $\mm$.
 Une section diff\'erentiable (continue) est une collection de sections locales diff\'erentiables
(continues) $\{\sigma_i\}$ telle que l'union des $U_i$ soit un
recouvrement de $\mm$. On note
$\Gamma^{\infty}(E)$\index{gammainfinie@$\Gamma^{\infty}(E)$}
(resp. $\Gamma(E)$)\index{gammae@$\Gamma(E)$} l'ensemble des
sections diff\'erentiables (continues) de $E$. C'est le module
(par convention, \`a droite) sur l'alg\`ebre $\cinf$ (resp.
$C(\mm)$) des fonctions lisses (continues) sur $\mm$,
\begin{equation}
\label{modulesection} (\sigma_1 + \sigma_2f)(x) \doteq \sigma_1(x)
+ \sigma_2(x)f(x)
\end{equation}
pour tout $\sigma_1, \sigma_2\in\Gamma^\infty(E)$. En prenant $F =
\rr^n$, o\`u $n$ est la dimension de $\mm$,  et $\pi^{-1}(x) = T_x
\mm$, on construit le fibr\'e vectoriel r\'eel
$TM$\index{tm@$TM$}, appel\'e {\it fibr\'e tangent}.  L'ensemble
des sections
$$\mathcal{X}(\mm) \doteq \Gamma^\infty(TM)$$
\index{xm@$\mathcal{X}(\mm)$} est l'ensemble des champs de
vecteurs lisses sur $\mm$. De mani\`ere analogue, on construit le
{\it fibr\'e cotangent} $T^*M$\index{tetoilem@$T^*M$} dont les
sections
$$\Omega^1(\mm)\doteq \Gamma^\infty(T^*M)$$
\index{omegaunm@$\Omega^1(\mm)$} sont les champs de $1$-forme.

Une {\it m\'etrique riemannienne} $g$\index{g} est une application
bilin\'eaire sym\'etrique ($g(X,Y) = g(Y,X)$), d\'efinie positive
($g(X,X)> 0$ pour $X\neq 0$) de $\cal{X}(\mm) \times \cal{X}(\mm)$
dans $\cinf$. Si $g$ est seulement non d\'eg\'en\'er\'ee ($g(X,Y)=
0 \text{ pour tout } Y \, \Rightarrow X = 0$), la m\'etrique est
dite {\it pseudo-riemannienne}.
 Dans les deux cas, $g$ d\'efinit une bijection de $\cinf$-module entre ${\cal X}(M)$ et $\Omega^1(M)$,
la bijection {\it musicale} $\flat\sharp$
\begin{eqnarray*}
{\cal{X}}(\mm)\rightarrow  \Omega^1(\mm)&:& X \mapsto \index{xbemol@$X^\flat$}X^\flat\; \text{ tel que }\, X^\flat(Y)\doteq g(X,Y),\\
\Omega^1(\mm)\rightarrow \cal{X}(\mm)&:&
\varpi\index{omegabar@$\varpi$} \mapsto \varpi^\sharp \; \text{
tel que }\, g(\varpi^\sharp,Y) \doteq \varpi(Y),
\end{eqnarray*}
o\`u $\varpi(.), X^\flat(.)$ d\'esignent l'action par dualit\'e de
$\Omega^1(\mm)$ sur $\xx(\mm)$. Le {\it gradient} d'une fonction
$f\in\cinf$ est par d\'efinition
\begin{equation}
\label{gradientdiese} \index{gradient}\grad{f} \doteq df^\sharp.
\end{equation}
 La m\'etrique induit
 une forme bilin\'eaire sym\'etrique d\'efinie positive (ou seulement non d\'eg\'en\'er\'ee dans le cas pseudo-riemannien) sur $\Omega^1(M)$, pareillement not\'ee $g$,
 \begin{equation}
\label{scldiese} g(\varpi_1, \varpi_2)\doteq g(\varpi_1^\sharp,
\varpi_2^\sharp).
\end{equation}
En tout $x$, $T_xM$ et $T_x^*M$ sont munis de la norme de
\begin{equation}
\label{normetangent} \norm{\grad{f}}\doteq g(\grad{\bar{f}},
\grad{f}) = g (d\bar{f}, df) \doteq \norm{df}.
\end{equation}

\section{Structure de spin}

\subsection*{Module de Clifford}

L'{\it alg\`ebre ext\'erieure} $\Lambda V$\index{lambdav@$\Lambda
V$} sur un espace vectoriel r\'eel $V$ est l'alg\`ebre formelle
g\'en\'er\'ee par un \'el\'ement identit\'e $\ii$ et  les
produits $v_1\wedge... \wedge v_k$ avec $v_1, v_k \in V$, $k\leq
\text{dim }V$, $v_1\wedge v_2 = - v_2\wedge v_1$ et $\ii\wedge v =
v$.  Lorsque $V$ est munie d'une forme bilin\'eaire non
d\'eg\'en\'er\'ee $g$, sym\'etrique \`a valeur dans $\rr$, on
construit l'{\it alg\`ebre de Clifford}
$\text{Cl}(V,g)$\index{clvg@$\text{Cl}(V,g)$} en "quantifiant" la
relation d'anticommutation de l'alg\`ebre ext\'erieure \`a l'aide
de $g$. Concr\`etement, $\text{ Cl}(V,g)$ en tant qu'espace
vectoriel est identique \`a $\Lambda V$ mais le produit est
d\'efini de sorte que
\begin{equation}
\label{gmunuclifford} uv + vu = 2 g(u,v)\ii
\end{equation}
 pour tout $u,v \in V$.
Avec $V^\cc \doteq V + iV$ le complexifi\'e de $V$ et l'extension
de $g$,  $g(u, v+iw) = g(u,v) + i g(u,w)$, on construit de la
m\^eme mani\`ere l'alg\`ebre de Clifford complexe $\cc \text{l}
(V)$. On omet $g$ dans la notation car toutes les formes non
d\'eg\'en\'er\'ees sur $V + iV$ donnent des alg\`ebres de Clifford
isomorphes. On obtient ainsi [\citelow{jgb}, {\it Lem. 5.5}]
\begin{equation}
\label{cliffordmatrice} \ccl(\rr^{2m}) \simeq M_{2^m}(\cc)\,
\text{ et } \ccl(\rr^{2m+1}) \simeq M_{2^m}(\cc) \oplus
M_{2^m}(\cc).
\end{equation}
$\ccl(V)$ est munie d'une involution $*$, obtenue en \'etendant
\begin{equation}
\label{clinv} (\lambda v_1... v_k)^* = \bar{\lambda} v_k... v_1
\end{equation}
 avec $\lambda\in\cc$, $v_1,..., v_k \in V,$ par lin\'earit\'e \`a tout $\ccl(V)$ (restreint \`a $V$ l'involution co\"{\i}ncide avec l'identit\'e, ce qui est
coh\'erent puisque $V$ est un espace vectoriel r\'eel).

Un \'el\'ement de $\ccl(V)$ est pair lorsqu'il s'\'ecrit comme
combinaison lin\'eaire de produits d'un nombre pair de vecteurs de
$V$. On note $\ccl ^+ (V)$ \index{clvplus@$\ccl ^+ (V)$} la
sous-alg\`ebre g\'en\'er\'ee par les \'el\'ements pairs, et
$\ccl^-(V)$\index{clvmoins@$\ccl^-(V)$} le sous-espace vectoriel
des produits impairs de vecteurs. En tant qu'espace vectoriel,
$\ccl (V) = \ccl^+(V) \oplus \ccl^- (V)$. On note
$\chi$\index{chi@$\chi$} la $\zz_2$ graduation correspondante
\begin{equation}
\label{defchi} \chi(a) = \pm 1 \text{ pour } a\in \ccl^{\pm}(V).
\end{equation}

Lorsque $g$ est d\'efinie positive, on d\'efinit l'\'el\'ement
{\it chiralit\'e} de $\ccl(V)$
\begin{equation}
\label{defgamma} \gamma\index{gamma@$\gamma$}\doteq (-i)^m
e_1e_2... e_n
\end{equation}
o\`u  $\{e_i\}$ est une base de $V$ orthonorm\'ee pour $g$ et
$n=\text{dim } V = 2m$ o\`u $2m+1$. Modulo l'orientation, $\gamma$
est ind\'ependant du choix de la base orthonorm\'ee. On v\'erifie
que $\gamma^2 = \gamma^*\gamma =\ii$. La chiralit\'e anticommute
ou commute avec $V$ selon que $n$ est pair ou impair. Lorsque $n$
est pair, $\gamma v \gamma = -v$ pour tout $v$ de $V$. Etendu \`a
tout $\ccl(V)$, on montre que $\gamma . \gamma$ co\"{\i}ncide avec
la $\zz_2$ graduation $\chi$. Lorsque $n$ est impair, $\gamma .
\gamma$ est l'identit\'e. La restriction, $g$ d\'efinie positive,
est fondamentale car c'est elle qui par la suite nous oblige \`a
consid\'erer des vari\'et\'es riemanniennes.
%
\newline

La m\'etrique $g$ d'une vari\'et\'e $\mm$ est d\'efinie sur les
sections lisses du fibr\'e tangent $TM$. Les champs de vecteurs
lisses sont denses dans l'ensemble
 des champs de vecteurs continus,
et $g$ s'\'etend en une forme bilin\'eaire sur les sections
continues $\Gamma(TM)$. Par complexification,
 on obtient une forme bilin\'eaire, encore not\'ee $g$, sur les sections continues du fibr\'e vectoriel complexe de fibre $T_x M^\cc = T_x M + i T_x M$\index{txmc@$T_xM^\cc$}.
Sur chacune de ces fibres $g$ induit une forme bilin\'eaire
permettant de former en tout $x$ de $\mm$ l'alg\`ebre de Clifford
$\ccl(T_x M)$.  Le fibr\'e vectoriel
 sur $\mm$ correspondant est not\'e $\ccl\, TM$. Le $C(\mm)$-module $\Gamma(\ccl\, TM)$
des sections continues de ce fibr\'e est une $C^*$-alg\`ebre,
produit et involution \'etant d\'efinis point par point
$$\sigma_1\sigma_2(x) \doteq \sigma_1(x)\sigma_2(x),\;\, \sigma^*(x)\doteq \sigma(x)^*\quad \forall x\in\mm\index{sigma@$\sigma$}$$
o\`u $*$ d\'esigne l'involution dans chaque $\ccl (T_xM)$, et la
norme est
$$\norm{\sigma}= \suup{x\in\mm} \{ \norm{\sigma(x)}\}$$
o\`u la norme de $\sigma(x)$ est celle de la $C^*$-alg\`ebre
$\ccl(T_xM)$. La construction est identique pour le fibr\'e
cotangent $T^*M$, ou pour n'importe quel fibr\'e vectoriel r\'eel
$E$ sur $\mm$, munie d'une forme bilin\'eaire non
d\'eg\'en\'er\'ee de $\Gamma^{\infty} (E) \times \Gamma^\infty
(E)$ dans $\cinf$. Pour disposer d'une chiralit\'e, on se limite
aux m\'etriques riemanniennes.

{\defi Le fibr\'e de Clifford sur une vari\'et\'e riemannienne $M$
de m\'etrique $g$ est le fibr\'e $\ccl(M)\doteq
\ccl\,T^*M$\index{clm@$\ccl(M)$}.}
\newline

Evalu\'ee en un point $x$ de $\mm$, une section $\sigma$ d'un
fibr\'e de Clifford est un \'el\'ement $\sigma(x)$ de
$\ccl(T^*_xM)$. Si $F$ est un espace vectoriel complexe sur lequel
agissent chacune des alg\`ebres $\ccl(T^*_xM)$ via {\it l'action
de Clifford}
\begin{equation}
\label{actionc} c\index{c}:\, \ccl(T^*_xM) \rightarrow
\text{End}(F),
\end{equation}
alors une section $\sigma$ du module de Clifford agit (par
convention \`a gauche) sur une section $\sigma'$ d'un fibr\'e
vectoriel $E\overset{\pi}{\rightarrow} \mm$ de fibre $F$ (i.e.
$\pi^{-1}(x)\simeq F$ pour tout $x$) par
$$\lp c(\sigma)\sigma'\rp(x) \doteq c(\sigma(x))\sigma'(x).$$
Lorsque l'action de $c(\sigma)$ est continue, c'est \`a dire
lorsque $c(\sigma)\sigma' \in \Gamma(E)$ pour tout
$\sigma\in\Gamma(\ccl(M)$ et $\sigma'\in\Gamma(E)$, $\Gamma(E)$
est un $\Gamma(\ccl(M))$-module \`a gauche. $\Gamma(E)$ \'etant
d\'ej\`a un $C(\mm)$-module \`a droite, c'est un bimodule.

{\defi Un module de Clifford sur $\mm$ est la donn\'ee d'un
$C(\mm)$-module $\Gamma(E)$ des sections continues d'un fibr\'e
vectoriel complexe sur $M$ ainsi que d'un homorphisme
$C(\mm)$-lin\'eaire
$$c:\, \Gamma(\ccl(\mm) ) \rightarrow \text{End}(\Gamma(E)).$$
Autrement dit, un module de Clifford sur $\mm$ est un
$\Gamma(\ccl(M))$-$C(\mm)$-bimodule de sections d'un fibr\'e
vectoriel complexe sur $\mm$.}
\newline

\noindent Si $\text{dim }\mm = 2m$, d'apr\`es
(\ref{cliffordmatrice}) toutes les actions irr\'eductibles de
$\ccl(\mm)$ sont de dimension $2^m$. Si $\text{dim }\mm = 2m+1$,
il y a deux repr\'esentations irr\'eductibles in\'equivalentes de
dimension $2^m$. Quand le rang du fibr\'e $E$ (i.e. la dimension
de ses fibres en tant qu'espace vectoriel) n'est pas \'egale \`a
$2^m$ dans le cas pair, $2^{m+1}$ dans
 le cas impair, chaque fibre $E_x$ se d\'ecompose en somme directe de sous-espaces vectoriels invariant par l'action de l'alg\`ebre de Clifford.
Au contraire quand l'action de l'alg\`ebre de Clifford est
irr\'eductible sur chaque fibre, $\Gamma(E)$ est un {\it module de
Clifford irr\'eductible}.

\subsection*{Groupe Spin}

Classiquement, le fibr\'e des spineurs sur une vari\'et\'e $\mm$
de dimension $n$ est construit \`a partir du fibr\'e tangent par
le rel\`evement du groupe $SO(n)$
 (groupe de structure du fibr\'e principal associ\'e au fibr\'e tangent)  \`a son recouvrement universel $\spin(n)$. L'approche alg\'ebrique construit
 directement un spineur comme support d'une action irr\'eductible du groupe $\spin$, vu
 comme sous groupe de l'alg\`ebre de Clifford.
 \newline

 Soit $V$ un espace vectoriel munie d'une forme bilin\'eaire non d\'eg\'en\'er\'ee $g$.
Un vecteur $u\in V$ est {\it unitaire} quand $g(u,u) = 1$. Vu
comme \'el\'ement de $\ccl(V)$,  $u^2 = \ii$ par
(\ref{gmunuclifford}) donc $u$ est inversible. On note $\phi(u)$
l'endomorphisme de $V$
$$\phi(u) v \doteq \chi(u) v u^{-1} = -u v u = (vu - 2 g(u,v))u = v- 2g(u,v)u$$
o\`u $\chi$ est la $\zz_2$ graduation d\'efinie en (\ref{defchi}).
Restreinte \`a $V$, qui est laiss\'e globalement invariant,
l'action de $\phi(u)$ est la r\'eflexion par rapport \`a
l'hyperplan orthogonal \`a $u$ (pour s'en convaincre on peut
regarder $\ccl(\rr^2)$ avec pour $g$ le produit scalaire usuel).
Par la multiplication
$$\phi_{u_1u_2}(v) \doteq u_2^{-1}u_1^{-1} v u_1 u_2 = \phi_{u_2}\circ\phi_{u_1}(v),$$
ces r\'eflexions g\'en\`erent le groupe orthogonal $O(V)$.
L'ensemble des produits pairs de r\'eflexions est le sous-groupe
des rotations $SO(V)$ (c'est la composante connexe de l'identit\'e
de $O(V)$). L'ensemble des produits pairs de vecteurs unitaires
$w$ de $V^\cc$ ($w = \lambda u$ avec $\lambda$ un nombre complexe
de module $1$ et $u$ un unitaire de $V$) est un sous groupe de
$\ccl(V)$ not\'e $\text{Spin}^c(V)$.

Pour tout $w\in \spin^c(V)$\index{spinc@$\spin^c$}, l'application
$\phi(w):\, v \mapsto w v w$ ($\chi(w) = +1$) est une rotation
dans $V$. $\phi$ apparait comme un homomorphisme de $\spin^c(V)$
dans $SO(V)$. Un \'el\'ement du noyau de $\phi$ est un unitaire
central pair de $\ccl(V)$ et on montre [\citelow{jgb}, {\it p.
180}] qu'un tel \'el\'ement est n\'ecessairement un scalaire.
Autrement dit $\text{ker }\phi\simeq U(1)$. Pour
$w=w_1...w_{2k}\in \spin^c(V)$, on d\'efinit l'homorphisme $\nu$
\`a  valeur dans $U(1)$
$$\nu(w) = w_{2k} ... w_1 w_1 ... w_{2k} = \lambda_1 ... \lambda_{2k}
$$
o\`u $\lambda_i = w_i^2\in U(1)$. Le groupe {\it
\spin(V)}\index{Spin} est par d\'efinition le noyau de $\nu$. La
conjugaison complexe est d\'efinie sur tout $\ccl(V)$ en
\'etendant par lin\'earit\'e $\overline{\lambda v} \doteq
\bar{\lambda} v$ pour $\lambda\in\cc$, $v\in V$. $\text{Spin}(V)$
est l'ensemble des unitaires pairs $w$ de $\ccl(V)$  satisfaisant
$\overline{w^*} w = w^* w = \ii$, ou encore $\bar{w} = w$. En
d\'efinissant la conjugaison de charge $\kappa$
$$\index{kappa@$\kappa$}\kappa(a) \doteq \chi(\bar{a}) $$
pour tout $a\in\ccl(V)$, le groupe Spin apparait comme le sous
groupe de $\spin^c$ invariant par conjugaison de charge. Le noyau
de $\phi$ restreint \`a $\spin(V)$ est $\{-1, 1\}$. En prenant
$V=\rr^n$ et $g$ une m\'etrique (pseudo-)riemannienne, on retrouve
que le spin est le recouvrement universel \`a deux feuillets du
groupe des rotations.

\subsection*{Vari\'et\'e \`a spin}

Par d\'efinition, un spineur est une section d'un fibr\'e
vectoriel sur une vari\'et\'e $M$ dont chaque fibre est le support
d'une repr\'esentation irr\'eductible du groupe $\spin(M)\doteq
\spin(T^*_xM)$. Le module des sections continues de ce fibr\'e est
donc un module de Clifford irr\'eductible. Cette propri\'et\'e
n'est pas suffisante pour caract\'eriser de mani\`ere alg\'ebrique
un fibr\'e de spineurs car rien ne garantit que tout module de
Clifford $\Gamma(E)$, fusse t'il irr\'eductible, soit le module
des sections d'un fibr\'e de spineurs. Dans le cas o\`u $M$ est de
dimension $n= 2m$ paire, on obtient un module de spineur en
demandant que $\Gamma(E)$ impl\'emente une {\it \'equivalence de
Morita} entre $C(M)$ et $\Gamma(\ccl(M))$.

{\defi\label{morita} Deux $C^*$-alg\`ebres $\aa$ et $\bb$ sont
Morita-\'equivalentes si et seulement si il existe un $\aa$-module
\`a droite plein $\ee$ tel que $\text{End}^{\, 0}_\aa(\ee) \simeq
\bb$, o\`u $\text{End}^{\, 0}_\aa(\ee)$ est la fermeture (pour la
topologie de la norme d'op\'erateur) de l'alg\`ebre des
endomorphismes de $\ee$ de $\aa$-rang fini.}
\newline

\noindent Cette d\'efinition demande plusieurs pr\'ecisions. Un
$\aa$-module $\ee$ est plein lorsqu'il est muni d'un "produit
scalaire \`a valeur dans $\aa$", c'est \`a dire d'une forme de
$\ee\times \ee$ dans $\aa$ d\'efinie positive, $\aa$-lin\'eaire
\`a droite, antisym\'etrique (\,$(u \lvert v) = (v \lvert u)^*$),
et telle que $(\ee \lvert \ee) = \aa$. Un endomorphisme de $\ee$
est dit de $\aa$-rang fini lorsqu'il est du type:
$$\lvert r) (s\lvert:\,t \longmapsto r(s\lvert t),$$
$r,s,t\in \ee$. Ces op\'erateurs forment une alg\`ebre qu'on munit
de la norme d'op\'erateur
$$\text{ sup}\{\norm{r(s\lvert t)}\, / \, \norm{t}= 1\}$$ o\`u la norme dans $\ee$ est
d\'efinie \`a partir de la norme de $\aa$ par $\norm{t}\doteq
\sqrt{\norm{(t\lvert t)}}.$

On montre que tout $\ee$ impl\'ementant l'\'equivalence de Morita
entre deux $C^*$-alg\`ebres $\aa$ et $\bb$ est n\'ecessairement de
type projectif fini en tant que $\aa$ module (cf condition 3,
section III.4 pour la d\'efinition d'un module projectif fini). Si
$\aa= C(M)$, alors d'apr\`es le th\'eor\`eme de Serre-Swan, $\ee$
est le module des sections continues d'un fibr\'e vectoriel sur
$M$: $\ee=\Gamma(E)$. Si de plus $\bb = \Gamma(\ccl(M))$, on
montre qu'il existe un isomorphisme de fibr\'e vectoriel
$\text{End } E\simeq \ccl(M)$ o\`u $\text{End } E$ d\'esigne le
fibr\'e vectoriel sur $M$ de fibre $\text{ End}(E_x)$. $\ccl(M)$
est de rang $2^n$, donc $\text{ End } E$ est de rang $2^n$, ce qui
signifie que $E_x$ est de dimension $\sqrt{2^n} = 2^m$. On peut
donc choisir l'action de Clifford de telle sorte que $\Gamma(E)$
soit un module de Clifford irr\'eductible. Rien n'assure en
revanche qu'impl\'ementer l'\'equivalence de Morita soit une
condition n\'ecessaire pour que $\Gamma(E)$ soit irr\'eductible.
Mais il apparait que la condition sur $M$ pour que $C(M)$ et
$\Gamma(\ccl(M)$ soit Morita \'equivalente (th\'eor\`eme de Plymen
[\citelow{jgb}, {\it Th. 9.3}]) est tr\`es exactement la condition
qui, dans l'approche classique,  autorise le rel\`evement de
$SO(n)$ au groupe $\spin^c(n)$.

La possibilit\'e du rel\`evement \`a $\spin(n)$  correspond [{\it
ibid, Th. 9.6}] \`a l'existence d'une bijection antilin\'eaire
$J:\, \Gamma(E) \rightarrow \Gamma(E)$ telle que
\begin{eqnarray}
\nonumber
\index{J} J(\psi f ) &=& (J\psi) \bar{f} \text{ pour } f\in C(M),\\
\label{bijc}
J(a \psi ) &=& \chi(a) J \psi  \text{ pour } a\in \ginf(\ccl(M)),\\
\nonumber ( J\phi \lvert J \psi ) &=& ( \psi \lvert \phi) \text{
pour } \phi, \psi \in \Gamma(E)
\end{eqnarray}
o\`u $\psi\in \Gamma(E)$ et on identifie $a$ et $f$ \`a leurs
actions sur $\Gamma(E)$. On montre [{\it ibid, Lem 9.7}] qu'un tel
op\'erateur $J$ est n\'ecessairement de carr\'e $\pm 1$.

Le produit scalaire sur $\Gamma(E)$ \`a valeur dans $C(M)$ est
choisi de sorte que l'action de Clifford soit autoadjointe,
$$( \phi  \lvert c(a)\psi ) = (c(a^*)\phi \lvert \psi).$$
On note $c(a)^\dagger = c(a^*).$ Si $M$ est {\it orient\'ee}, il
existe un rep\`ere mobile de $1$-formes $\{e_i\}$ (i.e. une
section lisse du fibr\'e cotangent) tel qu'en tout $x$ de $M$ les
chiralit\'es $\gamma(x)$ d\'efinies par (\ref{defgamma}) sur
chaque fibre de $\ccl(M)$ s'\'ecrivent
$$\gamma(x) = (-i)^m e_1(x)... e_n(x).$$
$\gamma$ est une section de $\ccl(M)$ et $c(\gamma)$ est une
graduation (i.e. $c(\gamma)^\dagger c(\gamma) = c(\gamma)^2 =
\ii)$ de $\Gamma(E)$. On note
$\Gamma(E)^{\pm}$\index{gammaplusmoins@$\gamma^{\pm}$} les
sous-espaces propres de $c(\gamma)$ de sorte que
$$
\Gamma(E) = \Gamma(E)^+ \oplus \Gamma(E)^-.
$$
Si $M$ est de dimension paire, $c(\gamma)$ anticommute avec
$c(\varpi)$ pour toute $1$-forme $\varpi\in\ccl(M)$. Pour
$\psi^{\pm}\in\Gamma(E)^{\pm}$,
$$c(\gamma) c(\varpi) \psi^{\pm} = \mp c(\varpi)\psi^\pm,$$
autrement dit $c(\gamma)$ \'echange $\Gamma^+(E)$ et
$\Gamma^-(E)$.

 {\defi Une {\it structure de spin} sur une vari\'et\'e $M$ de dimension paire est la donn\'ee d'un
 bimodule
\index{S} $S$ garantissant l'\'equivalence de Morita
 $C(M)$-$\Gamma(\ccl(M))$, d'une bijection $J$ satisfaisant (\ref{bijc}) et d'une orientation de $M$.}
\newline

\noindent $M$ est alors dite {\it vari\'et\'e \`a spin}. Lorsque
$M$ est de dimension impaire, la construction est analogue en
rempla\c{c}ant $\ccl(M)$ par $\ccl^+(M)$ qui est le fibr\'e sur
$M$ de fibre
  $\ccl^+(T_x^*M)$.

\section{Op\'erateur de Dirac}

\subsection*{Connexion}

Deux vecteurs en des points differents de $M$ appartiennent \`a
deux espaces tangents distincts, il n'y a donc a priori pas de
sens \`a vouloir les comparer. Pour r\'ealiser une telle
comparaison, on doit au pr\'ealalable d\'efinir la notion de {\it
transport parall\`ele}, c'est \`a dire la mani\`ere de transporter
un vecteur d'un point $x$ en un point $y$ "sans le modifier". La
premi\`ere id\'ee consiste simplement \`a d\'efinir le
transport\'e parall\`ele d'un vecteur $v^\mu \partial_\mu \in T_x
M$ comme le vecteur $v^\mu \partial_\mu' \in T_y M$ o\`u
$\partial_\mu$ est la base en coordonn\'ees locales de $T_xM$ et
$\partial_\mu'$ la base locale de $T_yM$. Si la vari\'et\'e est
munie de coordonn\'ees cart\'esiennes globales, c'est \`a dire si
$M\simeq\rr^m$, cette d\'efinition du transport parall\`ele
coincide avec la vision \'el\'ementaire du vecteur comme la
donn\'ee d'une "fl\`eche en un point". Ce n'est d\'ej\`a plus le
cas en coordonn\'ees polaires et c'est encore moins vrai quand la
vari\'et\'e ne dispose pas d'un syst\`eme de coordonn\'ees
globales. On d\'efinit donc le transport\'e parall\`ele d'un
vecteur $\partial_\nu$ de $T_x M$ dans la direction $dx^\mu$ comme
le vecteur
$$\triangledown_\mu (\partial_\nu) \doteq
\Gamma_{\mu\nu}^{\lambda} \partial_\lambda
$$
de $T_{x+dx^\mu} M$, o\`u $\Gamma_{\mu\nu}^{\lambda}\in\cc$ sont
appel\'es {\it coefficients de la connexion}. De mani\`ere
\'equivalente, une connexion associe \`a tout vecteur
$\partial_\mu$ de $T_x M$ une 1-forme de $T_{x + dx^\mu}M$ \`a
valeur vectoriel
$$\triangledown:\,  \partial_\mu \mapsto \Gamma_{\mu \alpha}^{\lambda}\partial_\lambda\otimes
dx^{\alpha}
 $$ o\`u
 $$\Gamma_{\mu \alpha}^{\lambda} \scl{dx^\alpha}{\partial_\nu} = \Gamma_{\mu
 \nu}^{\lambda}.
 $$
 Ainsi, une connexion sur un fibr\'e vectoriel $E\overset{\pi}\rightarrow M$ est une application lin\'eaire
 \begin{equation}
\label{connexion} \index{nabla@$\triangledown$}\triangledown:\,
\Gamma^\infty(E) \longrightarrow \Gamma^\infty(E)\ot \Omega^1(M)
\end{equation}
 satisfaisant la r\`egle de Leibniz
 $$\triangledown(\sigma f) = (\triangledown \sigma)f + \sigma\ot df$$
 pour tout $\sigma\in\Gamma^\infty(E)$ et $f\in\cinf.$ $d$ d\'esigne la d\'eriv\'ee ext\'erieure de chaque $\Lambda T^*_x M$ \'etendue aux sections lisses.
  Les {\it coefficients de connexion} sont obtenus en \'ecrivant localement, dans une base $\{dx^\mu\}$ de $\Omega^1(\mm)$, l'action de la connexion sur une base locale $\{\sigma_i\}$ de $\Gamma^\infty(E)$,
\begin{equation}
\label{coeffconnex} \triangledown \sigma_i \doteq \Gamma^{j}_
{i\mu}\index{gammajimu@$\Gamma^{j}_ {i\mu}$} \sigma_j \ot dx^\mu.
\end{equation}
Ainsi
\begin{eqnarray}
\nonumber
\triangledown \sigma = \triangledown \sigma_i f^i &=& (\triangledown \sigma_i)f^i + \sigma_i\ot d(f^i),\\
\nonumber
 &=& f^i\Gamma^{j}_ {i\mu} \sigma_j \ot dx^\mu  + \sigma_i\ot d(f^i),\\
\label{connexlocale} &=& (d + \Gamma) \sigma
 \end{eqnarray}
 o\`u on note $d \sigma \doteq \sigma_i\ot d(f^i) \;\text{ et }\; \Gamma \sigma \doteq f^i\Gamma^j_{i\mu} \sigma_j \ot dx^\mu.$

Lorsque $E$ est le fibr\'e tangent $TM$ sur une vari\'et\'e
riemannienne ou pseudo-riemannienne, il existe une unique
connexion, la connexion de {\it Levi-Civita}, de torsion nulle
(cf. [\citelow{nakahara}] pour une d\'efinition de la torsion) et
compatible avec la m\'etrique de la mani\`ere suivante:
\begin{equation}
\label{lvg} g(\triangledown X,Y) + g(X, \triangledown Y) = d
(g(X,Y))
\end{equation}
pour tout $X,Y\in\cal{X}(\mm)$. $g$ agit sur
$(\xx(\mm)\ot\Omega^1(\mm)) \times \xx(\mm)$ par "contraction des
indices"
$$g( r^i_\nu \partial_i\ot dx^\nu, t^\lambda \partial_\lambda) \doteq  r^i_\nu t^\lambda g(\partial_i, \partial_\lambda)dx^\nu,$$
$r^{i}_\nu, t^\lambda$ \'etant des nombres r\'eels. Apr\`es
identification de $g(X, .)$ \`a $X^\flat = \varpi$, (\ref{lvg})
d\'efinit une connexion de Levi-Civita sur $E= T^*M$ \`a valeur
dans $\Gamma^\infty(T^*M)\ot\Omega^1(\mm)\simeq
\Omega^1(M)\ot\Omega^1(M)$,
\begin{equation}
\label{lv1f} \triangledown \varpi (Y) \doteq d (\varpi(Y)) -
\varpi(\triangledown Y),
\end{equation}
pour tout $Y$ de $\xx(\mm)$. Pour $\varpi = dx^i$,
$Y=\partial_\mu$, (\ref{lv1f}) avec (\ref{coeffconnex}) donne
$\triangledown dx^i (\partial_\nu) = -  \Gamma^i_{\nu\mu} dx^\mu$
d'o\`u
\begin{equation*}
\label{connex1forme} \triangledown dx^i  = -  \Gamma^i_{j\mu} dx^j
\ot dx^\mu.
\end{equation*}
Localement l'action de $\del$ sur un $\varpi = dx^i f_i\in T^*M$
s'\'ecrit
$$\triangledown \varpi = (d - \tilde{\Gamma}) \varpi$$
o\`u $d \varpi \doteq dx^i\ot d(f_i)$ et $\tilde{\Gamma}\varpi
\doteq   - f_i \Gamma^i_{j\mu} dx^j \ot dx^\mu$. Par application
r\'ecursive de la r\`egle de Leibniz, $\del$ s'\'etend \`a tout
$\Gamma^\infty(\ccl(\mm)$,
\begin{equation}
\label{connexioncl} \triangledown (uv) \doteq \triangledown(u) v +
u \triangledown(v)
\end{equation}
pour tout $u,v\in \Gamma^\infty(\ccl(\mm))$ (le produit
$(\triangledown u)v$ consiste \`a multiplier les composantes dans
l'alg\`ebre de Clifford en laissant invariante la partie
$\Omega^1(\mm)$).

\subsection*{Connexion de spin}

Pour une vari\'et\'e \`a spin $(M, S, C)$, il existe une unique
{\it connexion de spin} $\del^S$
\index{nablas@$\triangledown^S$}g\'en\'eralisant la connexion de
Levi-Civita tout en \'etant compatible avec la structure de spin
[\citelow{jgb}, {\it Th. 9.8}].

{\thm Soit $(M, S, J)$ une vari\'et\'e \`a spin de dimension $n$.
Il existe une unique connexion $\del^S:\, \ginf(S) \rightarrow
\ginf(S)\ot \Omega^1(M)$ hermitienne, i.e.
$$( \del^S \psi \lvert \phi) + (\psi \lvert \del^S \phi) = d(\psi \lvert \phi) ,$$
commutant avec $J$ et telle que
$$\del^S( c(a)\psi) = c(\del a)\psi + c(a) \del^S\psi \text{ pour } a\in\ccl(M), \psi\in \ginf(S) $$
o\`u $c$ d\'esigne l'action de $\Gamma^\infty(\ccl(M))$ sur
$\ginf(S)$ induite par (\ref{actionc}) et $\del$ la connexion
(\ref{connexioncl}).}
\newline

\noindent L'action de l'alg\`ebre de Clifford se r\'e\'ecrit comme
une application de $\ginf(S)\ot \ginf(\ccl(M)) $ dans $\ginf(S)$
en posant
$$\hat{c}( \psi\ot a) \doteq c(a)\psi.$$

\subsubsection*{D\'efinition de l'op\'erateur de Dirac}

L'objet fondamental d'une g\'eom\'etrie spinorielle est l'{\it
op\'erateur de Dirac}, d\'efini comme suit.

{\defi \label{dirac} L'op\'erateur de Dirac d'une vari\'et\'e \`a
spin $(M, S, J)$ est l'endomorphisme de $\ginf(S)$
$$D\index{D}\doteq -i (\hat{c} \circ\del^S).$$}
\newline

Cet objet co\"{\i}ncide bien avec l'op\'erateur de Dirac de la
th\'eorie quantique des champs. Pour s'en convaincre, \'ecrivons
localement l'action de la connexion de spin. Tout espace de
Hilbert de dimension finie admettant
 une base orthonorm\'ee, il existe en tout $x$ de $M$ une base orthonorm\'ee de $T_xM$, $\{\partial_\alpha = e_\alpha^\mu(x) \partial_\mu\}$,
ainsi qu'une base duale de $T^*_xM$, \'egalement orthonorm\'ee,
$\{dx^\alpha = e^\alpha_\mu dx^\mu\}$. Le {\it vielbein}
$\{e^\alpha_\mu\}$ d\'esigne la matrice inverse de
$\{e_\alpha^\mu\}$ et satisfait
\begin{equation}
\label{geed} g^{\mu\nu} e^\alpha_\mu e^\beta_\nu =
\delta^{\alpha\beta}
\end{equation}
 o\`u $g^{\mu\nu} = g(dx^\mu, dx^\nu)$. Soit
$\{\gamma^a\index{gamma@$\gamma^a$}, \gamma^b\}$ un champ de
matrices de Dirac, i.e. des matrices autoadjointes de $M_k(\cc)$
($k= 2^{[n/2]}$ est la dimension de la repr\'esentation
irr\'eductible de $\ccl(M)$ dans le cas pair, de $\ccl(M)^+$ dans
le cas impair) telles que
\begin{equation}
\label{diraclide} \gamma^a(x) \gamma^b(x) + \gamma^b(x)
\gamma^a(x) = 2\delta^{ab}\ii
\end{equation}
en tout $x$ de $M$. En d\'efinissant la repr\'esentation
\begin{equation}
\label{cliffaction} c(dx^\alpha) \doteq \gamma^a,
\end{equation}
l'action de $dx^\mu\in\ccl(M)$ sur $S$
\begin{equation}
\label{matgamma} c(dx^\mu) \psi \doteq \gamma^m \psi \doteq
e^\mu_\alpha \gamma^a \psi
\end{equation}
(on utilise un indice grec pour les coordonn\'ees de la
vari\'et\'e et un indice latin pour le fibr\'e, $a$ est
contract\'e avec $\alpha$) est bien une repr\'esentation
(irr\'eductible) de $\ccl(M)$ puisque
\begin{eqnarray*}
c(dx^\mu)c(dx^\nu) + c(dx^\nu)c(dx^\mu) &=& 2e^\mu_\alpha e^\nu_\beta  \delta^{ab} \ii,\\
                                &=& 2 g^{\lambda\rho} e^\mu_\alpha e^\alpha_\lambda e^\nu_\beta   e^\beta_\rho \ii,\\
                                &=& 2g(dx^\mu, dx^\nu)\ii = c( dx^\mu dx^\nu + dx^\nu dx^\mu).
\end{eqnarray*}
On montre alors que la connexion de spin s'\'ecrit
\begin{equation}
\label{connecspin} \del^S  = d - \frac{1}{4}{\Gamma}^i_{j\mu}
\gamma_i \gamma^j\ot dx^\mu
\end{equation}
o\`u $\gamma_i\doteq \gamma^i$ et $d$ agit sur un spineur $\psi=
s_if^i$ \--$f^i\in C(M)$, $s_i\in\ginf(S)$\-- selon $d \psi \doteq
s_i \ot d(f^i)$. Quand la vari\'et\'e est plate  (ce qui est le
cas en th\'eorie des champs quand on suppose que l'interaction a
un lieu dans une r\'egion o\`u la courbure est localement
n\'egligeable) et que l'on choisit les coordonn\'ees
cart\'esiennes, les coefficients de connexion sont nuls et
\begin{equation}
\label{diracplat} iD\psi = \hat{c} (d\psi) = \hat{c} (s_i
\ot\partial_\alpha f^i dx^\alpha) = c(dx^\alpha) s_i
\partial_\alpha f^i  = \gamma^a s_i \partial_\alpha f^i = \ds
\psi.
\end{equation}

\section{Triplets spectraux: axiomes de la g\'eom\'etrie non
commutative}

Toute l'information g\'eom\'etrique d'une vari\'et\'e \`a spin, en
particulier la m\'etrique, est contenu dans l'op\'erateur de
Dirac. Cette remarque, dont nous rappelons dans cette section les
points cl\'es, est fondamentale puisqu'en donnant une d\'efinition
alg\'ebrique (i.e. en terme d'op\'erateur) des objets de la
g\'eom\'etrie spinorielle, elle permet de voir la vari\'et\'e \`a
spin commme un cas particulier, commutatif, d'une th\'eorie
beaucoup plus g\'en\'erale permettant de d\'efinir la
g\'eom\'etrie d'espaces non commutatifs. L'objet math\'ematique
d\'ecrivant ces g\'eom\'etries est le {\it triplet spectral
r\'eel}. Sa d\'efinition proc\`ede par \'etapes successives, en
commen\c{c}ant par isoler les propri\'et\'es essentielles (bien
s\^ur, elles n'apparaissent comme esssentielles qu'une fois la
construction achev\'ee)
 de l'op\'erateur de Dirac.

{\prop Si $D$ est l'op\'erateur de Dirac sur une vari\'et\'e \`a
spin $M$, alors
 \label{commutprop}
  $$
  [D, f] = -ic(df) \text{ pour tout } f\in\cinf.$$}

 \noindent{\it Preuve.} L'action de Clifford est $C(M)$-lin\'eaire, $c(af)\psi = c(a)\psi f$, donc
  \begin{equation*}
  i[D,f] \psi = \hat{c} (\del^S (\psi f)) - \hat{c}(\del^S\psi)f = \hat{c}\lp  \del^S (\psi f) - (\del^S\psi)f \rp = \hat{c} (\psi \ot df) = c(df)\psi,
  \end{equation*}
 pour tout $a\in\ginf(\ccl(M))$, $f\in\cinf$ et $\psi\in\ginf(S)$. \hfill $\blacksquare$
 \newline

 Gr\^ace au facteur $-i$ dans la d\'efinition \ref{dirac}, l'op\'erateur de Dirac est autoadjoint. $D$ \'etant non born\'e, cette affirmation
 n\'ecessite quelques pr\'ecautions. Notons tout d'abord qu'il existe un produit scalaire dans $\ginf(S)$,
 \begin{equation}
 \label{hilbspin}
 \scl{\psi}{\phi} \doteq \int_\mm (\psi \lvert \phi) \abs{\nu_g}
 \end{equation}
 o\`u
\begin{equation}
\label{nug} \index{nug@$\nu_g$} \nu_g = \sqrt{\text{det } g }\;
dx^1\wedge... \wedge  dx^n
\end{equation}
 est la forme volume de la vari\'et\'e $M$ et $g$ la matrice de composante $g(dx^\mu, dx^\nu)$.
 On renvoie aux ouvrages de g\'eom\'etries diff\'erentielles pour une \'etude
 de la th\'eorie de l'int\'egration sur une vari\'et\'e. Ici, il nous suffit de savoir que  (\ref{hilbspin}) co\"{\i}ncide localement avec l'int\'egrale de
 Lebesgue.  On note
 \begin{equation}
 \label{hld}
 \hh = L_2(M,S) \index{ldeuxms@$L_2(M,S)$}
 \end{equation}
 l'espace de Hilbert obtenue par compl\'etion de $\ginf(S)$ par rapport \`a la norme issue de ce produit scalaire. Dans le cas o\`u $M$ est plate, $L_2(M,S)$ est
l'espace des {\it spineurs de carr\'e sommable}
 de la m\'ecanique quantique. On conserve la m\^eme terminologie dans le cas g\'en\'eral. On montre alors que $D$ est formellement autoadjoint
 ($D=D^\dagger$) sur $\ginf(S)$, puis qu'il est essentiellement autoadjoint sur $\hh$, c'est \`a dire que
 $(D^\dagger)^\dagger$ est autoadjoint sur le sous-espace de $\hh$ compos\'e des spineurs $\psi_n$ pour lesquels \`a toute suite convergente $\psi_n\rightarrow \psi$
 correspond un spineur $\phi$ tel que $D\psi_n \rightarrow \phi$. Dans la suite, on identifie $D$ \`a $(D^\dagger)^\dagger$ en \'ecrivant simplement que $D$
 est autoadjoint.

 Quand $M$ est de dimension paire, la graduation $c(\gamma)$ anticommute avec $c(a)$ pour tout $a\in \Gamma^-(\ccl(M))$. Avec
(\ref{connecspin}), (\ref{diracplat}) et (\ref{matgamma}),
 $$ D \psi = -i c(\gamma) \lp \gamma^a \partial_\alpha  -  \frac{1}{4}\Gamma^{i}_{j\mu}
e^\mu_\alpha \gamma^a\gamma_i\gamma^j\rp \psi.$$
 Comme $\gamma^j$ et $\gamma^a\gamma_i\gamma^j$ appartiennent \`a $\Gamma^-(\ccl(M))$, $c(\gamma)$ anticommute avec l'op\'erateur de Dirac,
 \begin{equation}
 \label{chiralite}
 D\Gamma = - \Gamma D\index{Gamma@$\Gamma$}
 \end{equation}
 o\`u $\Gamma$ d\'esigne l'endomorphisme unitaire autoadjoint de $\hh$, extension de $c(\gamma)$, appel\'e {\it chiralit\'e}
 (on garde la m\^eme appellation pour $\gamma$ et $\Gamma$). A noter que $\Gamma$ est une $\zz_2$ graduation de $\hh$
 (pour \'eviter un conflit de notation, dans toute la suite $\Gamma$ d\'esigne la chiralit\'e et {\bf ne} d\'esigne {\bf plus} l'op\'erateur
 apparaissant dans la d\'efinition de la connexion de Levi-Civita).
\newline

 L'ensemble de ces propri\'et\'es est regroup\'e et g\'en\'eralis\'e dans les notions de {\it triplet spectral} et de $KR$-cycle. Pour tout
 $a\in\ginf(\ccl(M))$, $c(a)$ est un endomorphisme born\'e de $\hh$ car $c(a)$ est une collection d'op\'erateurs $c(a)_x$ agissant irr\'eductiblement sur
 les fibres $S_x$ de dimension $\text{dim }\ccl(T_xM^*)$ finie. Pour les m\^emes raisons, d'apr\`es la proposition  \ref{commutprop}, $[D,f]$ est born\'e,
 de m\^eme que $f$ qui agit simplement par multiplication sur $\hh$.

 {\defi \label{tripletspectral} Un triplet spectral $(\aa, \hh, D)$ pour une alg\`ebre $\aa$ est la donn\'ee d'un espace de Hilbert $\hh$, d'une repr\'esentation
 de $\aa$ dans l'alg\`ebre $\bb(\hh)$  des op\'erateurs born\'es sur $\hh$, et d'un op\'erateur autoadjoint $D$, de r\'esolvante compacte, tel
 que $[D,a]\in\bb(\hh)$ pour tout $a\in\aa$.}
\newline

\noindent Rappelons qu'un op\'erateur $D$ est \`a r\'esolvante
compacte\cite{reed} si et seulement si pour tout $\lambda\notin
\text{sp}(D)$, $(D- \lambda\ii)^{-1}$ est compact (un op\'erateur
$T$ sur $\hh$ est compact quand, pour $\epsilon>0$, $\norm{T}\leq
\epsilon$ sauf sur un sous-espace de $\hh$ de dimension fini).

 Lorsque $\aa$ est une alg\`ebre involutive, on d\'efinit la notion de {$KR$-cycle}.

 {\defi \label{krcycle}
 Soit $n\in\zz_8$; un $KR^n$-cycle pour une alg\`ebre involutive $\aa$ est un triplet spectral $(\aa, \hh, D)$ accompagn\'e de
 \begin{itemize}
 \item une bijection unitaire $J$ antilin\'eaire sur $\hh$ qui impl\'emente l'involution, i.e. $ J a J^{-1} = a^*$ pour tout $a$ de $\aa$;
 \item si $n$ est pair, une graduation $\Gamma$ de $\hh$ qui commute avec $\aa$ et anticommute avec $D$;
 \item la table de multiplication-commutation suivante
  \begin{center}
   \begin{tabular}{|c|cccccccc|}
   \hline
   n mod 8&0&1&2&3&4&5&6&7\\
   \hline
   $J^2 = \pm \ii$&+&+&\--&-&-&-&+&+\\
   $JD= \pm DJ$&+&-&+&+&+&-&+&+\\
   $J\Gamma = \pm \Gamma J$&+& &-& &+& &-&\\
   \hline
   \end{tabular}
 \end{center}
\end{itemize}
\noindent Pour $n$ impair, on pose $\Gamma =\ii$ (qui
naturellement commute avec $D$ et $J$) et on note de fa\c{c}on
g\'en\'erale $(\aa, \hh, D, \Gamma, J)$ un $KR$-cycle.}
\newline

Si $(M, S, J)$ est une vari\'et\'e \`a spin de dimension $n$, $D$
l'op\'erateur de Dirac et $\Gamma$ l'extension de $c(\gamma)$ aux
spineurs de carr\'e sommable, alors  $(\cinf, L_2(M,S), D, J,
\Gamma)$ est un $KR^n$-cycle. Que $(\cinf, \hh, D)$ soit un
triplet spectral est \'evident compte tenu de la discussion
pr\'ec\'edent  la d\'efinition \ref{tripletspectral} (on renvoie
\`a [\citelow{connes,jgb}] pour prouver que $D$ est \`a
r\'esolvante compacte); que $J$ impl\'emente l'involution (la
conjugaison complexe) de $\cinf$ d\'ecoule de (\ref{bijc});
l'action de Clifford est $C(M)$-lin\'eaire donc $\Gamma$ commute
avec la repr\'esentation de $\aa$;  l'anticommutation de $D$ et
$\Gamma$ est \'etablie en (\ref{chiralite}); reste la table de
commutation, montr\'ee en d\'etail dans [\citelow{jgb}, {\it Th.
9.19}]. A toute vari\'et\'e \`a spin est associ\'e un $KR$-cycle
mais l'inverse n'est pas vrai: la donn\'ee d'un $KR$-cycle ne
suffit pas
 \`a construire une vari\'et\'e \`a spin. Pour ce faire, il faut ajouter une s\'erie de conditions d\'etaill\'ees ci-dessous.
 \newline

  Nous donnons directement les conditions pour qu'un triplet spectral $(\aa, \hh, D)$ d\'efinisse une {\it g\'eom\'etrie non commutative}\cite{gravity}, en
  rappelant ensuite comment, adapt\'ees au cas commutatif, ces conditions tiennent lieu d'axiomes d'une vari\'et\'e \`a spin. Les trois premi\`eres
conditions sont plus analytiques qu'alg\'ebriques. Elles sont
importantes dans la d\'efinition axiomatique de la g\'eom\'etrie
commutative spinorielle mais dans les exemples
   \'etudi\'es dans cette th\`ese (g\'eom\'etrie de dimension z\'ero ou produit de g\'eom\'etries dont l'une est commutative)
   elles sont toujours remplies. Nous les donnons ici par exhaustivit\'e, en renvoyant \`a [\citelow{connes,gravity,jgb}] pour une d\'efinition pr\'ecise des objets
   qu'elles font intervenir.

{\cond[Dimension]  L'op\'erateur $D^{-1}$ est un infinit\'esimal
d'ordre $\frac{1}{n}$ o\`u $n\in\nn$ est la dimension (spectrale)
de la g\'eom\'etrie.}
\newline

\noindent $D$ \'etant \`a r\'esolvante compacte, $D^{-1}$
d\'efinit en restreignant $D$ \`a $\hh/\ker D$ est un op\'erateur
compact. Ainsi\cite{reed} la suite
 d\'ecroissante $\{\lambda_k\}$ des valeurs propres de $\abs{D^{-1}} \doteq  \sqrt{(D^{-1})^\dagger D^{-1}}$ tend vers z\'ero. $D^{-1}$ est un infinit\'esimal
 d'ordre $\frac{1}{n}$  signifie que cette suite d\'ecroit au moins aussi vite que $k^{-n}$,
 $$\lim_{k\rightarrow +\infty} \lambda_k = O(\frac{1}{k^n}).$$
Lorsque $\aa$ et $\hh$ sont de dimension finie, la dimension
spectrale est nulle.

{\cond[R\'egularit\'e] Pour tout $a\in\aa$, $a$ et $[D,a]$
appartiennent \`a l'intersection des domaines de toutes les
puissances $\delta^k$ de la d\'erivation $\delta(b)\doteq [
\abs{D}, b]$, o\`u  $b$ est \'el\'ement de l'alg\`ebre
g\'en\'er\'ee par $\aa$ et $[D, \aa]$.}
\newline

\noindent Cette condition est la version alg\'ebrique de la
diff\'erentiabilit\'e des coordonn\'ees.

{\cond[Finitude] $\aa$ est une pr\'e-$C^*$-alg\`ebre  et
l'ensemble $\index{hinfini@$\hh^{\infty}$}\hh^\infty\doteq
\underset{k\in \nn} \cap \text{ Dom } D^k$ des vecteurs lisses de
$\hh$ est un module projectif fini.}
\newline

\noindent Un $\aa$-module est libre quand il a une base $\{e_i\}$,
c'est \`a dire un ensemble de g\'en\'erateurs tels que $a^i e_i =
0$ pour $a^i\in\aa$ implique $a^i = 0$ pour tout $i$. Un module
{\it projectif} est une somme directe de modules libres. Un tel
module n'est pas forc\'ement libre. Il est {\it fini} lorsqu'il a
une famille g\'en\'eratrice de cardinalit\'e finie.  Une
pr\'e-$C^*$-alg\`ebre est une sous alg\`ebre d'une
$C^*$-alg\`ebre, stable par le calcul fonctionnelle holomorphe (cf
[\citelow{jgb},  {\it Def. 3.26}]). En particulier $\cinf$ est une
pr\'e-$C^*$-alg\`ebre.
\newline

Les quatre conditions restantes sont d'ordre alg\'ebriques et ce
sont elles qui seront discut\'ees dans les mod\`eles des chapitres
suivants. Au triplet spectral $(\aa, \hh, D)$ est adjoint une {\it
chiralit\'e} $\Gamma$ c'est \`a dire, lorsque la dimension
spectrale $n$ est paire, une $\zz_2$ graduation $\Gamma=\Gamma^2$
de $\hh$, autoadjointe, qui anticommute avec $D$ et commute avec
la repr\'esentation de $\aa$. On note $\hh^{\pm}$ les sous-espaces
propres de $\Gamma$. $D$ envoie un sous-espace dense de
$\hh^{\pm}$ dans $\hh^{\mp}$ si bien que, dans la d\'ecomposition
$\hh = \hh^+ + \hh^-$,
\begin{equation}
\label{dpm}D = \dm{cc} 0& D^-\\ D^+ & 0\fm\index{dplus@$D^+, D^-$}
\end{equation}
 o\`u
\begin{equation}
\label{dplus} D^+ \doteq \frac{\ii - \Gamma}{2}D\frac{\ii +
\Gamma}{2}
\end{equation}
 et $D^- = (D^+)^\dagger$. Quand $n$ est impair, $\Gamma=\ii$.

On demande \'egalement que $\hh$ soit le support d'une
repr\'esentation de l'alg\`ebre oppos\'ee $\aa^{\circ}$ (identique
\`a $\aa$ en tant qu'espace vectoriel mais o\`u le produit est
invers\'e: $a^{\circ}b^{\circ} = (ba)^{\circ}$)
 impl\'ement\'ee par un op\'erateur unitaire antilin\'eaire  $J$,
$$b^{\circ} \mapsto Jb^*J^{-1},$$
tel que
\begin{equation}
\label{commutj} [a, Jb^* J^{-1}] =0.
\end{equation}
La repr\'esentation de $\aa^{\circ}$ commute avec la
repr\'esentation de $\aa$, et $\hh$ porte une repr\'esentation
$\pi$ de l'alg\`ebre involutive $\aa\ot\aa^{\circ}$
$$a\ot b^{\circ} \mapsto aJb^*J^{-1}$$
o\`u l'involution est donn\'e par $(a\ot b^{\circ})^* \doteq b^*
\ot (a^*)^\circ.$ De mani\`ere \'equivalente, on dit que $\aa$ est
repr\'esent\'ee
 \`a gauche et $\aa^{\circ}$ \`a droite
$$a\psi b \doteq aJb^*J^{-1} \psi = Jb^*J^{-1}a \psi.$$
 Si $J^2= \pm \ii$, $J b^* J^{-1} = J^{-1} b^* J$ de sorte que
\begin{eqnarray*}
\pi ( (a\ot b^\circ)^*) &=& b^* J a J^{-1} = J a J^{-1} b^*,\\
                        &=& J (a J^{-1} b^* J) J^{-1} = J (a J b^* J^{-1})J^{-1},\\
                        &=& J \pi (a\ot b^\circ) J^{-1}
\end{eqnarray*}
et $J$ impl\'emente l'involution de $\aa\ot \aa^\circ$.

{\cond[R\'ealit\'e] $(\aa\ot\aa^{\circ}, \hh, D, \Gamma, J)$ est
un $KR^n$-cycle.  $J$ est appel\'ee {\it structure r\'eelle}.}

{\cond[Premier ordre] La repr\'esentation de $\aa^{\circ}$ commute
avec $[D,\aa]$
$$[[D,a], Jb^*J^{-1}] = 0 \,\text{ pour tout } \, a,b\in\aa.$$}
Cette condition stipule que l'op\'erateur de Dirac est un
op\'erateur diff\'erentiel du premier ordre.

{\cond[Orientabilit\'e] Il existe un cycle de Hochschild $c\in
Z_n(\aa, \aa\ot \aa^{\circ})$ tel que $\pi(c) = \Gamma$.}
\newline

\noindent Cette condition est la g\'en\'eralisation de la non
d\'eg\'en\'erescence de la forme volume pour une vari\'et\'e
orient\'ee. Avant de d\'efinir l'homologie de Hochschild,
rappelons qu'un {\it complexe} est une suite de $\aa$-module
$E^i$, $i\in\nn$, et de morphismes $d_i$ de $E_i$ dans $E_{i+1}$,
\begin{equation}
\label{complexe} ... \rightarrow E_{i-1}
\overset{d_{i-1}}\rightarrow E_{i} \overset{d_i}\rightarrow
E_{i+1} \rightarrow ...
\end{equation}
tels que $d_i\circ d_{i-1} = 0$. L'image d'un morphisme est
incluse dans le noyau du morphisme suivant;
 quand cette inclusion est une \'egalit\'e, $\text{Im } d_{i-1} = \ker d_i$, le complexe est {\it exact}. Sinon on note $Z_i\index{zi@$Z_i$}\doteq
 \ker d_i$ le module des {\it i-cycles} et $B_i\index{bi@$B^i$}\doteq \text{ Im }d_{i-1}$ le module des {\it i-bords}. Le quotient
 $H_i\doteq Z_i/B_i$ est par d\'efinition le {\it $i^{\text{\`eme}}$ groupe d'homologie} du complexe ($H_i$ est en fait un $\aa$-module). L'ensemble des $H_i$
  forme l'homologie du complexe.
 En rempla\c{c}ant $E_i$ par $C_i(\aa, N)\doteq N\ot\aa\ot...\ot\aa$ o\`u $N$ est un bimodule sur $\aa$ et le produit tensoriel de $\aa$ par elle-m\^eme est
 r\'ep\'et\'e $i$ fois, on a
$$... \overset{b}{\rightarrow} C_i(\aa, N) \overset{b}{\rightarrow} C_{i-1}(\aa, N) \rightarrow ... \overset{b}\rightarrow C_0(\aa, N) \overset{b}\rightarrow
  \{ 0 \}$$
  o\`u l'application $b$ de $C_i(\aa, N)$ dans $C_{i-1}(\aa,N)$ d\'efinie par
$$b(n\ot a_1\ot...\ot a_i) \doteq na_1 \ot a_2 \ot ... \ot a_i + \sum_{p=1}^{i-1} (-1)în\ot a_1 \ot...\ot a_pa_{p+1}\ot ... \ot a_i + (-1)^i a_in \ot a_1\ot
  ... \ot a_{i-1}$$
  v\'erifie $b^2 = 0$, on d\'efinit {\it l'homologie de Hochschild de $\aa$ \`a valeur dans le bimodule
$N$}. On munit $\aa\ot \aa^{\circ}$ d'une structure de
$\aa$-bimodule
  $$x(a\ot b^{\circ})y \doteq xay\ot b^{\circ} \, \text{ pour tout }\, x,y,a,b\in\aa$$
  de mani\`ere \`a d\'efinir l'homologie de Hochschild de $\aa$ \`a valeur dans $N = \aa\ot \aa^{\circ}$. Un {\it cycle de Hochschild} $c\in
  Z_n(\aa, \aa\ot \aa^{\circ})$ est un \'el\'ement de $C_n(\aa, \aa\ot \aa^{\circ})$
  tel que $b(c) =0$. Un \'el\'ement $c= a\ot b^{\circ}\ot a_1\ot...\ot a_n$ de $C_n(\aa, \aa\ot\aa^{\circ})$ est repr\'esent\'e sur $\hh$ par
  \begin{equation}
\label{pic} \pi(c) \doteq aJb^*J^{-1}[D, a_1]...[D,a_n].
\end{equation}
 Cette repr\'esentation est coh\'erente avec la proposition \ref{commutprop} qui identifie $[D,f]$ \`a la $1$-forme $df$, $1$-bord dans la cohomologie de de Rham (cf. ci-dessous). Elle est \'etendue \`a tout $C_n(\aa,
\aa\ot\aa^{\circ})$ par addition.

{\cond[Dualit\'e de Poincar\'e] Le couplage additif sur $K_*(\aa)
$ d\'etermin\'e par l'indice de l'op\'erateur de Dirac est
non-d\'eg\'en\'er\'e.}
\newline

\noindent Cette condition est une version alg\'ebrique de la
dualit\'e de Poincar\'e. Rappelons qu'en rempla\c{c}ant dans
(\ref{complexe}) $E_i$ par l'espace vectoriel r\'eel $\Omega^i(M)$
des $i$-formes sur une vari\'et\'e compacte $M$ de dimension $n$,
et $d^i$ par la diff\'erentielle ext\'erieure $d$, on d\'efinit un
complexe dont l'homologie  est appel\'ee {\it cohomologie de de
Rham}. La dualit\'e de Poincar\'e stipule que pour tout entier
positif $r\leq n$, les groupes de cohomologie de de Rham $H^r(M)$
et $H^{n-r}(M)$ sont duaux; c'est \`a dire qu'il existe une forme
bilin\'eaire non-d\'eg\'en\'er\'ee de $H^r(M)\times H^{n-r}(M)$
dans $\rr$
$$\scl{[\varpi^r]}{[\varpi^{n-r}]} \doteq \int_M \varpi^r \wedge \varpi^{n-r}$$
o\`u $[\varpi^r]$ d\'esigne la classe d'\'equivalence dans
$H^r(M)$ de $\varpi^r\in\Omega^r(M)$. Gr\^ace au {\it caract\`ere
de Chern}, cette forme bilin\'eaire se traduit par un couplage
additif (i.e. une forme bi-additive) des groupes de $K$-th\'eorie
de l'alg\`ebre $\cinf$. Pour une d\'efinition de ces objets, on
peut consulter [\citelow{wegge}]. Ici, contentons nous de
souligner que ce couplage, not\'e $\cap$\index{inter@$\cap$},
s'effectue gr\^ace \`a l'indice de l'op\'erateur de Dirac
(d\'efini ci-dessous) et ne fait pas r\'ef\'erence \a la
commutativit\'e de l'alg\`ebre, de sorte qu'en rempla\c{c}ant
$\cinf$ par une pr\'e-$C^*$-alg\`ebre quelconque (pour de tels
objets, la $K$-th\'eorie existe et est identique \`a la
$K$-th\'eorie de la $C^*$-alg\'ebre obtenue par compl\'etion), la
condition 7 appara\^{\i}t comme la d\'efinition abstraite de la
dualit\'e de Poincar\'e.

Sans entrer dans le cas g\'en\'eral, pr\'ecisons un exemple qui
sera utile pour l'\'etude des espaces non commutatifs finis
(chapitre 3). Quand la dimension spectrale $n$ est paire, la
dualit\'e de Poincar\'e pour $r$ pair se ram\`ene au couplage
additif de $K_0(\aa)\index{kzeroa@$K_0(\aa)$}\times K_0(\aa)$ \`a
valeur dans $\zz$ (car pour tout $r\leq n$ pair,  $K_r(\aa)\simeq
K_{n-r}(\aa)\simeq K_0(\aa)$) d\'efini de la mani\`ere suivante.
On note $P_l(\aa)$\index{pla@$P_l(\aa)$} l'ensemble des
projecteurs de $M_l(\aa)$\index{mla@$M_l(\aa)$} (alg\`ebre des
matrices $l\times l$ \`a coefficients dans $\aa$) et $GL_l(\aa)$
les \'el\'ements inversibles de $M_l(\aa)$. On a les plongements
\'evidents
$$m\in M_l(\aa) \mapsto \dm{cc} m & 0 \\ 0 & 0 \fm \in M_{l+1}(\aa) \; \text{ et }\; v\in GL_l(\aa) \mapsto \dm{cc} v & 0 \\ 0 & 1\fm \in GL_{l+1}(\aa)$$
et on d\'efinit
$$M_\infty(\aa)\doteq \underset{l=1}{\overset{\infty}{\bigcup}} M_l(\aa)\, ,\;P_\infty(\aa)\doteq \underset{l=1}{\overset{\infty}{\bigcup}} P_l(\aa)\, , \;
GL_\infty(\aa)\doteq \underset{l=1}{\overset{\infty}{\bigcup}}
GL_l(\aa).$$ Deux projecteurs $p,q\in P_l(\aa)$ sont
\'equivalents, $p\sim q$ si et seulement si ils sont conjugu\'es
via un $v\in GL_{\infty}(\aa)$, c'est \`a dire s'il existe
$k\in\nn$ et $v\in GL_{k+l}(\aa)$ tels que
\begin{equation}
\label{equivpoinc} v\dm{cc} p & 0 \\ 0 & 0_k\fm v^{-1} = \dm{cc} q
& 0 \\ 0 & 0_k \fm.
\end{equation}
Le quotient $K_0^+(\aa)\doteq P_{\infty}(\aa) / \sim$ est un
semi-groupe (ie. les \'el\'ements ne sont pas inversibles) pour
l'addition
$$[p] + [q] \doteq \left[ \dm{cc} p& 0 \\ 0& q\fm\right] = \left[ \dm{cc} q& 0 \\ 0& p\fm\right].$$
$K_0(\aa)$ est par d\'efinition le groupe de
Grothendieck\cite{lang} de $K_0^+(\aa)$ dont les \'el\'ements sont
les classes d'\'equivalence de $K_0^+(\aa)\times K_0^+(\aa)
/\sim$, o\`u $(p,q)\sim(p',q')$ si et seulement si $p+q' = q +
p'$. $K_0(\aa)$ est un groupe pour l'addition $(p,q) + (p', q')
\doteq (p+p', q+q')$ avec l'\'el\'ement nul $(0,0)$ et l'inverse
$-(q,p)\doteq (p,q)$. C'est le m\^eme proc\'ed\'e qui permet de
construire $\zz$ \`a partir de $\nn$: $(p,q)$ s'identifiant \`a
$p-q$, on utilise la notation $\pm p$, $\pm q$ pour d\'esigner les
\'el\'ements de $K_0(\aa)$. Si $p\in M_k(\aa)$ et $q\in M_l(\aa)$,
alors $P\doteq p\ot (J\ot \ii_l)q(J^{-1}\ot\ii_l)$ est un
projecteur agissant sur $\hh\ot \cc^{kl}$ ($\aa$ est suppos\'ee
complexe) et
\begin{equation}
\label{intersection} \cap ([p], [q] ) \doteq \text{ indice} \lp
P(D\ot\ii_{kl})P\rp \doteq \dim\lp\ker P(D^+\ot\ii_{kl})P\rp -
\dim\lp\ker P(D^-\ot\ii_{kl})P\rp
\end{equation}
o\`u $D^+$, $D^-$ sont d\'efinis dans (\ref{dpm}).
\newline

{\defi \label{gnc} Un triplet spectral satisfaisant les sept
conditions ci-dessus est un {\it triplet spectral r\'eel}, ou
encore {\it une g\'eom\'etrie non commutative (spinorielle)},
not\'e $(\aa, \hh, D, \Gamma, J)$.}

\section{G\'eom\'etrie commutative}

On appelle {\it g\'eom\'etrie commutative} une g\'eom\'etrie non
commutative au sens de la d\'efinition \ref{gnc} o\`u l'alg\`ebre
$\aa$ est commutative. Parmi les g\'eom\'etries commutatives, les
{\it g\'eom\'etries de Dirac} sont les triplets spectraux r\'eels
\begin{equation}
\label{td} T_D = (\cinf, L_2(M,S), D, J, \Gamma)
\end{equation}
dans lesquels $(M, S, J)$ est une vari\'et\'e riemannienne
compacte orient\'ee \`a spin, $D$ est l'op\'erateur de Dirac
d\'efini par la connexion de spin et $\Gamma$ la chiralit\'e
(\ref{chiralite}) si $\text{dim }M = n$ est paire, $\Gamma=\ii$ si
$n$ est impair. Les g\'eom\'etries de Dirac sont bien des
g\'eom\'etries non  commutatives, c'est \`a dire que $T_D$
v\'erifient les $7$ conditions de la section pr\'ec\'edente. On
renvoie \`a [\citelow{jgb}] pour la preuve de cette affirmation.
Contentons nous de rappeler que la dimension spectrale de la
g\'eom\'etrie de Dirac est \'egale \`a la dimension de la
vari\'et\'e. Il est \'egalement int\'eressant de s'attarder sur la
condition d'orientabilit\'e dont l'appellation trouve son origine
dans les g\'eom\'etries de Dirac.

Notons tout d'abord les simplifications dues \`a la
commutativit\'e de l'alg\`ebre.  $\cinf$ est identique \`a son
alg\`ebre oppos\'ee. $L_2(M,S)$ est donc le support de deux
repr\'esentations distinctes, la multiplication \`a gauche par une
fonction $f$, et l'action \`a droite correspondant \`a la
multiplication par la fonction complexe conjugu\'ee $\bar{f}$.
Ainsi $JfJ^{-1}\psi = \bar{f}\psi$, ce qui est coh\'erent avec
(\ref{bijc}) puisque $J f \psi = JfJ^{-1}J\psi  = \bar{f}J{\psi}$
(\`a noter le changement de convention lors du passage du
$\cinf$-module droit $S$ \`a l'espace de Hilbert $L_2(M,S)$
$\cinf$-lin\'eaire \`a gauche). Dans (\ref{pic}), identifier
$J\bar{f}J^{-1}$ \`a $f$ permet de voir le cycle de Hochschild $c$
comme un \'el\'ement de $Z_n(\aa)$, ensemble des $n-cycles$ dans
l'homologie de Hochschild du complexe
$$... \overset{b}{\rightarrow} C_i(\aa) \overset{b}{\rightarrow} C_{i-1}(\aa) \rightarrow ... \overset{b}\rightarrow C_1(\aa) \overset{b}\rightarrow
  \{ 0 \}$$
o\`u $C_i(\aa)\doteq \aa \ot ... \ot \aa$ ($\aa$ apparait $i$
fois), repr\'esent\'e par
\begin{equation}
\label{piccom} \pi( f_0 \ot f_1\ot f_i) \doteq
f_0[D,f_1]\,...\,[D,f_i].
\end{equation}

Soit $\{U_j, x_j\}$ un atlas de $M$. $x_j$ est une fonction de
$U_j \rightarrow \rr^n$ et chacune de ses composantes $x_j^\mu$
est \'el\'ement de $C^{\infty}(U_j)$. La forme volume (\ref{nug})
est l'unique $n$-forme qui, \'evalu\'ee sur toute base
orthonorm\'ee et orient\'ee de $TM$, vaille $1$. Localement,
 $$\nu_g = \sqrt{\text{det }{g}}\; dx_j^1 \wedge ... \wedge dx_j^n$$
o\`u $g$ est la matrice de composante $g(\partial_\mu,
\partial_\nu)$ et (en omettant l'indice $j$) $\{\partial_\mu\}$
est la base locale de $TM$. La forme volume est ind\'ependante du
choix des coordonn\'ees sur l'ouvert $U_j$. Lorsque $\{
\theta_j^\alpha = e_{j\mu}^\alpha dx_j^\mu\}$
 est une base locale orthonorm\'ee de $1$-formes, $g$ est la matrice identit\'e et
$$\nu_g = \theta_j^1 \wedge ... \wedge \theta_j^n = e_j dx_j^1 \wedge ... \wedge dx_j^n$$
o\`u $e_j\doteq \text{det }\left\{  e_{j\mu}^\alpha \right\}$. On
pose $e^0_j\doteq i^{n-m}f_j e_j\in C^{\infty}(U_j)$, o\`u
$m\doteq [n/2]$ et $f$ est une {\it partition de l'unit\'e}, c'est
\`a dire un ensemble $\{f_j\in C^{\infty}(U_j)\}$ de fonctions
telles que
$$0\leq f_j(x) \leq 1,\quad f_j(x) = 0 \, \text{ pour } x\notin U_j,\quad \sum_j f_j(x) = 1 \text{ pour tout } x\in M.
$$
On en d\'eduit l'\'ecriture non-locale de l'\'el\'ement de volume,
$$i^{n-m}\nu_g = \sum_j e^0_j dx_j^1 \wedge... \wedge dx_j^n.$$
Le $n$-cycle de Hochschild $c\in Z_n(\aa)$ correspondant est, par
d\'efinition,
$$c\doteq \frac{1}{n!} \sum_{\sigma\in S_n} (-1)^\sigma \sum_j e^0_j\ot x_j^{\sigma(1)}\ot ... \ot x_j^{\sigma(n)},$$
o\`u $S_n$ est le groupe des permutations de $\{1,..., n\}$ et
$\sigma$ d\'esigne \`a la fois un \'el\'ement de $S_n$ et sa
parit\'e (l'exposant $\sigma$ \'egale $\pm 1$ selon que la
permutation $\sigma$ est paire ou impaire).  Par (\ref{piccom}),
en utilisant la proposition \ref{commutprop}, on v\'erifie que
\begin{eqnarray*}
\pi(c) &=& \frac{1}{n!} \sum_{\sigma\in S_n} (-1)^\sigma \sum_j e_j^0 [D, x_j^{\sigma(1)}]\, ... \, [D, x_j^{\sigma(n)}]\\
       &=& \frac{(-i)^n}{n!} \sum_{\sigma\in S_n} (-1)^\sigma \sum_j e_j^0  c(d x_j^{\sigma(1)})\, ... \, c(dx_j^{\sigma(n)})\\
       &=& \frac{(-i)^m}{n!} \sum_j f_j \sum_{\sigma\in S_n} (-1)^\sigma c(\theta_j^{\sigma(1)})\, ... \, c(\theta_j^{\sigma(n)})\\
       &=& (-i)^m \sum_j f_j  c(\theta_j^{1})\, ... \, c(\theta_j^{n}) \\
       &=& c(\gamma) \sum_j f_j = \Gamma.
\end{eqnarray*}
Ainsi la chiralit\'e  est bien l'image du cycle de Hochschild
correspondant \`a l'\'el\'ement de volume.
\newline

Si $\aa$, $D$, $J$ et $\Gamma$ commutent avec un projecteur $p$ de
$\hh$, alors la g\'eom\'etrie non commutative $(\aa, \hh, D, J,
\Gamma)$ peut s'\'ecrire comme somme directe de deux
g\'eom\'etries non commutatives d\'efinies sur $p\aa$ et
$(\ii-p)\aa$ (pour des g\'eom\'etries de Dirac, ceci correspond
\`a une vari\'et\'e non connexe). Pour \'eviter ces cas, on dit
qu'une g\'eom\'etrie non commutative  est {\it irr\'eductible}
lorsqu'il n'y a pas de projecteur non nul commutant avec $\aa$,
$D$, $J$ et $\Gamma$. Ainsi toute vari\'et\'e \`a spin connexe de
dimension $n$ d\'efinit par (\ref{td}) une g\'eom\'etrie de Dirac
irr\'eductible, c'est \`a dire une g\'eom\'etrie non commutative
irr\'eductible, de dimension spectrale $n$, d\'efinie sur
l'alg\`ebre $\cinf$. A l'inverse, selon le th\'eor\`eme suivant
\'enonc\'e dans [\citelow{gravity}] et dont on trouve une
d\'emonstration d\'etaill\'ee dans [\citelow{jgb}], toute
g\'eom\'etrie non commutative sur $\cinf$, irr\'eductible et de
dimension spectrale $n$ est une g\'eom\'etrie de Dirac pour une
vari\'et\'e \`a spin.

{\thm \label{thconnes}Soit $T=(\aa, \hh, D, J, \Gamma)$ une
g\'eom\'etrie non commutative irr\'eductible sur $\aa = \cinf$, de
dimension spectrale $n =\text{dim } M$ o\`u $M$ est une
vari\'et\'e compacte  orient\'ee connexe sans bord. Alors
\begin{itemize}
\item Il existe une unique m\'etrique riemannienne $g=g(D)$ sur $M$ telle que la distance g\'eod\'esique sur $M$ soit donn\'ee par
\begin{equation}
\label{distancecommut} d(x,y) =\suup{f\in C(M)} \{ f(x) - f(y)\, /
\, \norm{[D, f]}\leq 1 \}.
\end{equation}
\item $M$ est une vari\'et\'e \`a spin et les op\'erateurs $D'$ pour lesquelles $g(D') = g(D)$ forment une union d'espaces affines identifi\'es par les
structures de spin sur $M$.
\item La fonctionnelle $S(D)\doteq {\int\!\!\!\!\!\!{\--}} \abs{D}^{-n+2}$ d\'efinit une forme quadratique sur chacun de ces espaces affines, atteignant
son minimum pour $D = D_s$, l'op\'erateur de Dirac correspondant
\`a la structure de spin; ce minimum est proportionnel \`a
l'action d'Einstein-Hilbert, c'est \`a dire l'int\'egrale de la
courbure scalaire $s$
$$S(D_s) = -\frac{n-2}{24}\int_M s\sqrt{\text{det }g}\,d^nx.$$
\end{itemize}}

\noindent Les deux et troisi\`eme points n\'ecessitent quelques
explications. La structure de spin de $M$ est donn\'ee par le
bimodule $S = \hh^{\infty}$ d\'efini par l'op\'erateur $D$ (cf.
condition de finitude) et l'op\'erateur $J$. En r\`egle
g\'en\'erale, l'op\'erateur de Dirac $D_s$ correspondant \`a cette
structure de spin n'est pas l'op\'erateur $D$. La seule chose
qu'on puisse affirmer est que
$$D = D_s + \rho$$
pour $\rho\in\text{End}(\ginf(S))$ v\'erifiant
\begin{equation}
\label{condrho} \rho^\dagger = \rho,\; \Gamma\rho = (-1)^n \rho
\Gamma,\; J\rho J^{-1} = \pm \rho,
\end{equation}
le signe \'etant n\'egatif lorsque et seulement lorsque $n=1$ ou
$5$ mod $8$. Tout $\rho'$ satisfaisant (\ref{condrho}) d\'efinit
un op\'erateur $D'\doteq D_s + \rho'$ tel que $g(D') = g(D)$.
Ainsi pour la structure de spin donn\'ee par $T$, l'ensemble des
op\'erateurs d\'eterminant la m\^eme m\'etrique que $D$ est un
espace affine. Si maintenant on consid\`ere une vari\'et\'e
riemannienne $M$ o\`u la  m\'etrique $g$ est fix\'ee, il existe
plusieurs structures de spin sur $M$ (le nombre de structure de
spin est fini et est d\'etermin\'e par la cohomologie  de Cech de
$M$). Fixer une structure de spin d\'etermine de mani\`ere unique
l'op\'erateur $D_s$, et  les op\'erateur $D'$ du type $D_s + \rho$
d\'efinissent un ensemble de g\'eom\'etries \'equivalentes. Ainsi
les structures de spin d'une vari\'et\'e riemannienne $M$
permettent de classifier, au regard de la topologie,  les
g\'eom\'etries sur $M$.

La fonctionnelle du troisi\`eme point est d\'efinie par
$${\int\!\!\!\!\!\!{\--}} \abs{D}^{-n+2} \doteq \frac{1}{2^{[n/2]}\Omega_n} \text{Wres } \abs{D}^{-n+2}
$$
o\`u $\Omega_n$ est l'int\'egrale de la forme volume sur la
sph\`ere $S^n$ et $\text{ Wres}$ est le {\it r\'esidu de Wodzicki}
(cf [\citelow{jgb}, {\it Th. 7.5}] pour une d\'efinition).
\newline

Le triplet spectral r\'eel est un outil permettant de classifier
les g\'eom\'etries spinorielles sur une vari\'et\'e compacte (sans
bord). L'avantage de cette formulation alg\'ebrique est que la
d\'efinition \ref{gnc} est valable pour des pr\'e-$C^*$-alg\`ebres
quelconques, pas forc\'ement commutatives.  Dans la premi\`ere
partie de ce chapitre, on a vu que les $C^*$-alg\`ebre non
commutatives \'etaient des candidats s\'erieux pour jouer le
r\^ole de fonctions sur un espace non commutatif. De m\^eme que le
th\'eor\`eme de Gelfand, par analogie avec le cas commutatif,
justifie le choix des \'etats purs d'une $C^*$-alg\`ebre comme
points d'un espace non commutatif, de m\^eme le th\'eor\`eme
\ref{thconnes} sugg\`ere que le triplet spectral r\'eel est un bon
outil pour faire la g\'eom\'etrie de ces espaces non commutatifs.
En particulier, et c'est l'objet de cette th\`ese, la formule
(\ref{distancecommut}) dans sa
 formulation g\'en\'erale d\'efinit une distance sur l'espace des \'etats d'une alg\`ebre.

\chapter{La distance}

\section{La formule de la distance}

Classiquement, la distance entre deux points $x$, $y$ est la
longueur du plus court chemin reliant $x$ \`a $y$. Physiquement
cette mani\`ere de voir n'est pas acceptable
 car la m\'ecanique quantique invalide l'id\'ee d'un chemin entre deux points. D'autre part un point n'est pas accessible \`a l'exp\'erience autrement que par l'interm\'ediaire
 d'une observable. Pour concilier g\'eom\'etrie et m\'ecanique quantique, il faudrait donc d\'efinir une distance $d(x,y)$ qui ne fasse r\'ef\'erence qu'aux
 valeurs prises par les observables sur $x$ et $y$. Qu'apparaissent des valeurs d'observables sur d'autres points $p$ est tol\'er\'e,
 \`a condition que les-dits points soient caract\'eris\'es autrement que par une appartenance \`a un chemin entre $x$ et $y$. Par ailleurs une distance est par d\'efinition
 une fonction de deux variables \`a valeur r\'eelle, positive, sym\'etrique, r\'eflexive ($d(x,x)=0$) et qui  v\'erifie l'in\'egalit\'e triangulaire.
 La mani\`ere la plus simple d'impl\'ementer ces propri\'et\'es au niveau des observables est de consid\'erer une quantit\'e du type
$\abs{f(x) - f(y)}$ o\`u $f$ est une fonction complexe  sur
l'espace. Dans le cas le plus simple de la droite r\'eelle,
$d(x,y)=\abs{x-y}$. Pour que $\abs{f(x) - f(y)} = \abs{x-y}$ il
faut au moins que
\begin{equation}\label{principe}
\abs{f'(p)} =1 \text{ pour un point $p$ du segment $[x,y]$,}
\end{equation}
$f'$ d\'esignant la d\'eriv\'ee de $f$. Caract\'eriser $p$ par son
appartenance au segment $[x,y]$ viole les principes d'une "bonne"
distance au sens quantique. Heureusement la condition
(\ref{principe}) peut s'exprimer ind\'epen\-damment de $x$ et $y$.
En posant
\begin{equation}
\label{distreelle} d(x,y) = \suup{f\in C^1(\rr)} \left\{ \abs{f(x)
- f(y)}\; / \, \abs{f'(p)}\leq 1\, ,\forall p\in \rr\, \right\}
\end{equation}
o\`u $C^1(\rr)$ est l'ensemble des fonctions d\'erivables sur
$\rr$, on v\'erifie ais\'ement que $d(x,y)= \abs{x-y}$, le
supr\'emum \'etant atteint par la fonction de d\'eriv\'ee
constante (\'egale \`a $1$): $x\mapsto x$.

Dans cette formule les points demeurent les objets premiers (non
seulement parce qu'il s'agit d'une distance entre points, mais
aussi parce que sont privil\'egi\'ees  des valeurs d'observables
en des points pr\'ecis). Cependant le th\'eor\`eme de Gelfand
assure qu'un points $x$ de l'espace
 n'est rien d'autre qu'un \'etat pur $\omega_x$ de l'alg\`ebre commutative des observables continues sur cet espace. En repr\'esentant $\aa =
C^1(\rr)$ sur l'espace des fonctions r\'eelles de carr\'e sommable
$\hh = L_2(\rr)$ par simple multiplication point par point,
(\ref{distreelle}) s'\'ecrit
\begin{equation}
\label{distcorse} d(x,y) = \suup{f\in\aa} \left\{ \abs{\omega_x(f)
- \omega_y(f)}\, / \, \norm{\left[\frac{d}{dx}, f\right]} \leq 1
\right\}
\end{equation}
o\`u  $\frac{d}{dx}$ est l'op\'erateur de d\'erivation sur
$L_2(\rr)$. Pour s'en convaincre, il suffit de remarquer que pour
tout $\psi\in\hh$,
$$[\frac{d}{dx}, f]\psi = \frac{d}{dx} f\psi - f\frac{d}{dx}\psi = -\lp \frac{d}{dx}f\rp \psi = -f'\psi$$
d'o\`u
$$\norm{\left[\frac{d}{dx}, f\right]}= \suup{\psi\in\hh} \frac{\norm{f'\psi}}{\norm{\psi}} = \suup{\psi\in\hh}
\lp \frac{\int_\rr \abs{f'(x)}^2\abs{\psi(x)}^2 dx}{\int_\rr
\abs{\psi(x)}^2dx}\rp^{\frac{1}{2}} = \suup{x\in
\rr}\abs{f'(x)}.$$ En rempla\c{c}ant $\rr$ par une vari\'et\'e
riemannienne \`a spin $M$, $C^1(\rr)$ par $\aa = C(M)$, $L_2(\rr)$
par $\hh= L_2(M,S)$ et $\frac{d}{dx}$ par un op\'erateur $D$ tel
que $(\aa, \hh, D)$ soit un triplet spectral au sens de la
d\'efinition \ref{tripletspectral}, (\ref{distcorse}) est
identique \`a (\ref{distancecommut}). Cette d\'efinition de la
distance g\'eod\'esique, en apparence plus complexe que la
d\'efinition usuelle, est en fait plus pr\'ecise puisqu'elle se
g\'en\'eralise imm\'ediatement \`a tout triplet
spectral.\cite{metrique,connes}

{\defi Soit $(\aa, \hh, D)$ un triplet spectral. La distance $d$
entre deux \'etats $\tau_1$ et $\tau_2$  est
\begin{equation}
\label{distance} d(\tau_1,\tau_2)\doteq \sup_{a\in\aa} \left\{ \,
\abs{\tau_1(a) -\tau_2(a)}\, /\, \norm{[D,a]}\leq 1 \right\}.
\end{equation}}
On v\'erifie imm\'ediatement que cette distance est positive,
sym\'etrique
  et reflexive, et presque imm\'ediate\-ment
qu'elle satisfait l'in\'egalit\'e triangulaire puisque
\begin{eqnarray*}
d(\tau_1, \tau_2) &=& \sup_{a\in\aa} \left\{ \, \abs{\tau_1(a) -\tau_2(a)}\, /\, \norm{[D,a]}\leq 1 \right\},\\
                    &\leq&  \sup_{a\in\aa} \left\{ \,\abs{\tau_1(a) - \tau_3(a)} + \abs{\tau_3(a) -\tau_2(a)}\, /\, \norm{[D,a]}\leq 1\right\},\\
                    &\leq&  \sup_{a\in\aa} \left\{ \,\abs{\tau_1(a) - \tau_3(a)}\, /\, \norm{[D,a]}\leq 1 \right\} + \sup_{a\in\aa} \left\{ \,
\abs{\tau_1(a) - \tau_3(a)}\, /\, \norm{[D,a]}\leq 1\right\},\\
    &\leq& d(\tau_1, \tau_3) + d(\tau_3, \tau_2).
\end{eqnarray*}

A noter que cette d\'efinition n'impose pas au triplet spectral
d'\^etre r\'eel, la seule condition indispensable est que le
commutateur $[D,a]$ reste born\'e pour tout $a$. Dans le chapitre
suivant, on \'etudiera des exemples de distance associ\'ee \`a des
triplets r\'eels et \`a d'autres non r\'eels. De m\^eme $\aa$
n'est pas n\'ecessairement une (pr\'e)-$C^*$-alg\`ebre. Cependant
les propri\'et\'es des $C^*$-alg\`ebres, et \`a plus forte raison
celles des $W^*$-alg\`ebres, permettent de mener bon nombre de
calculs \`a terme. De plus c'est ce type d'alg\`ebre qui
s'interpr\`ete comme fonction sur l'espace non commutatif, on
s'int\'eresse donc dans la suite essentiellement aux triplets
spectraux sur des $C^*$-alg\`ebres.

Dans le cas commutatif, soulignons que (\ref{distancecommut}), qui
fait intervenir l'alg\`ebre des fonctions continues, n'est pas la
traduction exacte de (\ref{distance}) appliqu\'ee au triplet
(\ref{td}) construit sur l'alg\`ebre des fonctions lisses. La
formulation de (\ref{distancecommut}) est emprunt\'ee \`a
[\citelow{jgb}] qui reprend [\citelow{connes}] o\`u cette formule
est donn\'ee avec pour alg\`ebre $\aa$ l'alg\`ebre des fonctions
born\'ees mesurables sur $M$ (dense dans $C(M)$). Dans
[\citelow{gravity}], la formule de la distance est donn\'ee
directement pour $\aa=\cinf$.
\newline

Pour terminer cette section, citons un lemme qui g\'en\'eralise
l'id\'ee que la distance pour la droite r\'eelle est
"r\'ealis\'ee" par une fonction positive de d\'eriv\'ee partout
\'egale \`a $1$, (cf ref.[\citelow{these}] pour la preuves)

{\lem \label{positif} Si $d({\tau_1} , \tau_2)$ est finie alors
$d({\tau_1} , \tau_2) = \underset {a \in \mathcal{A}_{+}} {\sup}
\, \{  \, \abs{\tau_1(a) - \tau_2(a)}\;\, / \;\, \norm{[D,a]} = 1
\},$}
\newline
o\`u $\mathcal{A}_{+}$ d\'esigne l'ensemble des \'el\'ements
positifs de $\aa$ (cf page 7).

\section{Distance g\'eod\'esique pour une vari\'et\'e
riemanienne}

Le triplet spectral associ\'e \`a une vari\'et\'e riemannienne
compacte \`a spin avec une m\'etrique $g$ est donn\'e par
(\ref{td}).  D'ap\`es la proposition \ref{commutprop}, la partie
connexion de spin de l'op\'erateur de Dirac commute avec $\cinf$
si bien que
 pour tout $f\in\cinf$,
\begin{eqnarray*}
[D,f] &=& - i c(df) = -i \partial_\mu f c( dx^\mu) =  -i \partial_\mu f c( e^\mu_\alpha dx^\alpha),\\
      &=&  -i \partial_\mu f  e^\mu_\alpha\gamma^a = -i \gamma^m\partial_\mu f,\\
      &=& [-i\gamma^m\partial_\mu, f]
\end{eqnarray*}
o\`u $c$ est l'action de Clifford (\ref{cliffaction}),
$e^\alpha_\mu$ les vielbein, $\gamma^a$ les matrices de Dirac
euclidiennes (\ref{diraclide}) et $\gamma^m \doteq e^\mu_\alpha
\gamma^a$ les matrices de Dirac riemanniennes (\ref{matgamma}) qui
v\'erifient, gr\^ace \`a (\ref{geed}),
$$
\gamma^m \gamma^n + \gamma^n \gamma^m = 2 g^{\mu\nu}\ii$$
(conform\'ement au chapitre I, on utilise un indice grec pour la
vari\'et\'e et un indice latin pour les degr\'es de libert\'e de
spin; $a,m,n$ se contractent avec $\alpha, \mu, \nu$).   Pour le
calcul des distances, on peut consid\'erer que le triplet spectral
d'une vari\'et\'e \`a spin est
\begin{equation}
\label{triplettvariete} \aa=\cm,\qquad \hh=\LS,\qquad
D=-i\gamma^{m}\partial_{\mu}=-i\ds.
\end{equation}
La dimension spectrale est la dimension de la vari\'et\'e qu'on
prend \'egale \`a  $4$. Le triplet spectral est pair donc la
chiralit\'e (\ref{chiralite}) s'\'ecrit
$$\Gamma = c(\gamma) = (-i)^2 c( \overset{4}{\underset{\alpha=1}{\Pi}}dx^\alpha )  =
-\overset{4}{\underset{\alpha=1}{\Pi}}\gamma^a = -\gamma^5
$$
o\`u $\gamma$ est donn\'ee en (\ref{defgamma}) et $\gamma^5$
d\'esigne traditionnellement le produit des matrices gamma
euclidiennes. Le produit scalaire  de $\hh$ est donn\'e par
(\ref{hilbspin}) ou le couplage $(.\lvert.)$ de $S$ \`a valeur
dans $C(M)$ est le produit scalaire euclidien des spineurs vus
comme vecteurs colonnes dont les entr\'ees sont des fonctions
d'onde,
$$(\psi\lvert\phi) = \psi^\dagger \phi$$
o\`u  $\psi^\dagger$ d\'esigne le vecteur ligne complexe
conjugu\'e de $\psi$.

Comme \'enonc\'e dans le th\'eor\`eme \ref{thconnes}, la distance
non commutative (\ref{distance})
\begin{equation}
\label{distance1,5} d(x,y)=\sup_{f\in C(M)} \left\{ \, \abs{f(x)
-f(y)}\, /\, \norm{[D,f]}\leq 1 \right\}\,,
\end{equation}
coincide avec la distance g\'eod\'esique $L(x,y)$ entre les points
$x,y$ de $\mm$. C'est un r\'esultat classique \cite{connes} dont
nous rappelons la preuve:

Par la d\'efinition (\ref{clinv}) de l'involution dans $\ccl(M)$,
$(dx^\alpha)^* = dx^\alpha$ si bien qu'en choisissant les matrices
de Dirac autoadjointes, on choisit en fait une action de Clifford
autoadjointe,
$$(\gamma^a)^\dagger = \gamma^a = c(dx^\alpha) =c({dx^\alpha}^*).$$
Ainsi la norme d'op\'erateur de $[D,f]$, pour une fonction $f$
r\'eelle selon le lemme \ref{positif}, s'\'ecrit
\begin{eqnarray*}
\norm{[D,f]}^2 &=& \displaystyle \norm{c(df)^2} = \sup_{\psi \in
\hh}
\frac{ \int_{\mm} ( c(df)\psi \lvert c(df)\psi)\abs{\nu_g}}{\int_{\mm}(\psi\lvert \psi)\abs{\nu_g}},\\
&=& \displaystyle \sup_{\psi \in \hh} \frac{\int_M \psi^\dagger c(df^*) c(df)\psi \abs{\nu_g}}{\int_{\mm}\psi^\dagger \psi\abs{\nu_g}},\\
&=& \displaystyle \sup_{x\in\mm} \left\{
g^{\mu\nu}(x)\partial_{\mu}f(x)\partial_{\nu}f(x)\right\} =
\suup{x\in M} g(df, df) = \norm{\grad{f}}^2
 \end{eqnarray*}
o\`u on utilise $c(df^*)c(df) = \partial_\mu {f} \partial_\nu f
\gamma^m\gamma^n =  g^{\mu\nu} \partial_\mu f \partial_\nu f \ii$
ainsi que l'\'equation (\ref{normetangent}).

D'o\`u
$$
\norm{[D,f]}=\suup{x\in\mm}\norm{(\grad{f})(x)}.
$$

Soit maintenant  $c\!:\!t\in [0,1]\!\!\rightarrow\!\!\mm$ une
g\'eod\'esique minimale entre  $x$ et $y$. On d\'esigne par un
point la d\'eriv\'ee totale par rapport au param\`etre $t$.  Pour
tout $f\in\cinf$
$$f(x)-f(y) = \int_{0}^{1} \dot{f}(c(t))\; dt=\int_{0}^{1}
\partial_{\mu}f(p)\; \dot{c^{\mu}}(t) dt$$
avec $p\doteq c(t)$. Les fonctions $\dot{c^{\mu}}$ sont les
composantes d'un champ de vecteur $X\in\xx(M)$. On note
$\dot{c}_\nu$ les composantes de $\dot{c}\doteq X^\flat$, si bien
que
$$
{\partial_{\mu}f(p)\, \dot{c^{\mu}}(t)}={g^{\mu\nu}(p)\,
\partial_{\mu}f(p) \, \dot{c_{\nu}}(t)}= g( df(p), \dot{c}(p)).
$$
Par l'in\'egalit\'e de Cauchy-Schwarz,
$$\abs{{\partial_{\mu}f(p)\,\dot{c}^{\mu}(t)}} \leq \norm { df(p)} \norm{\dot{c}(t)}.$$
Si $f$ atteint le supr\'emum,
$\norm{df(p)}=\norm{(\grad{f})(p)}\leq1$ en tout $p$ et
\begin{equation*}
\label{firstineq} d(x,y) =\abs{f(x)-f(y)}\leq \int_0^1
\norm{\dot{c}(t)}\; dt = L(x,y)\,.
\end{equation*}

Cette borne sup\'erieure est atteinte par la fonction
\begin{equation}
\label{defl} \index{L@$L$}L: q \mapsto L(q,y).
\end{equation}
En effet, $L(x)-L(y) = L(x,y)$ et
\begin{equation}
\label{normedel} \sup_{q\in\mm}\norm{{\grad{L}}(q)}\leq 1\,.
\end{equation}
Pour montrer cette derni\`ere in\'egalit\'e,  choisissons
$q,q'\in\mm$, de coordonn\'ees $q^{\mu},{q'}^\mu$ dans une carte
donn\'ee, o\`u $q'$ est l'image de $q$ par la transformation
infinit\'esimale  $\sigma(\epsilon)$, $\epsilon<\!\!< 1$, $\sigma$
d\'esignant le flot g\'en\'er\'e par le champ de vecteurs
$g^{\mu\nu}(\partial_{\nu} L)\partial_{\mu}$ avec la condition
initiale $\sigma(0)=q$. Alors, avec $dq^{\mu}{\doteq
q'}^{\mu}-q^{\mu}$,
$$
q^{\mu}+dq^{\mu}= {q'}^{\mu} = \sigma^{\mu}(\epsilon) =
\sigma^{\mu}(0) + \epsilon\, \frac{d \sigma^{\mu}}{dt}(0) + {\cal
O}(\epsilon^2) = q^{\mu} +\epsilon\, g^{\mu\nu}(q)\partial_{\nu}
L(q) +{\cal O}(\epsilon^2)\,,
$$
c'est \`a dire
\begin{equation}
\label{flot} dq^{\mu}=\epsilon\,g^{\mu\nu}(q)\partial_{\nu} L(q)
+{\cal O}(\epsilon^2)\,.
\end{equation}
Comme $L(q',y)$ est le plus court chemin de $q'$ \`a $y$,
$L(q',y)\leq L(q',q) + L(q,y)$. Ainsi
\begin{equation}
\label{inegtrg} L(q+dq) \leq L(q',q) + L(q)\,.
\end{equation}
Par (\ref{flot})
$$
L(q',q)\doteq \sqrt{g_{\lambda\rho}(q) dq^{\lambda}dq^{\rho}}=
\sqrt{\epsilon^2 g_{\lambda\rho}(q)  g^{\lambda\mu}(q)
\partial_{\mu}L(q)\, g^{\rho\nu}(q) \partial_{\nu} L (q)}=
\epsilon \sqrt{ g^{\mu\nu}\partial_{\mu}L(q)\, \partial_{\nu} L
(q)}\,.
$$
Inser\'e dans le membre de droite de (\ref{inegtrg}) dont la
partie gauche est d\'evelopp\'ee par rapport \`a $\epsilon$, cette
\'equation donne
$$
L(q) + \partial_{\mu}  L(q) \, dq^{\mu}= L(q) + \,
\epsilon\,g^{\mu\nu}(q)\partial_{\mu}L(q)\partial_{\nu} L(q)
+{\cal O}(\epsilon^2) \leq
\epsilon\sqrt{g^{\mu\nu}\partial_{\mu}L(q)\,  \partial_{\nu} L
(q)} + L(q) +{\cal O}(\epsilon^2),
$$
qui est vraie quel que soit $q$, d'o\`u (\ref{normedel}) et
finalement $$d(x,y)=L(x,y).$$

A noter que $L$ n'est pas lisse en $y$ mais seulement
continue\cite{jgb}. Pour \'ecrire (\ref{distance1,5}) en
rempla\c{c}ant $C(M)$ par $\cinf$, il faudrait exhiber une suite
de fonctions lisses $f_n$ qui converge vers $L$ et v\'erifie
$\norm{[D, f_n]}\leq 1$ pour tout $n$.

\section{Exemples d'espaces finis}

Les exemples les plus simples d'espaces non commutatifs sont
associ\'es \`a des alg\`ebres de dimension finie.  On peut
r\'esoudre de mani\`ere syst\'ematique les contraintes impos\'ees
par les axiomes de la g\'eom\'etrie non commutative et \'etablir
une classification compl\`ete des triplets  spectraux
finis\cite{krajew,paschke}. Pr\'ecisons encore une fois que la
formule (\ref{distance}) d\'efinit une distance sur l'ensemble des
\'etats d'une \alg ind\'ependamment des axiomes de la
g\'eom\'etrie non commutative, aussi dans un premier temps nous
consid\'erons des triplets spectraux $(\aa, \hh, D)$ qui ne sont
pas r\'eels. En dimension fini, $D$ et $[D,a]$ pour tout $a\in
\aa$ sont born\'es. La seule contrainte qu'on impose \`a
l'op\'erateur de Dirac est d'\^etre autoadjoint. S'appuyant sur
l'\'equivalence dans le cas commutatif entre caract\`eres et
\'etats purs, on a choisi de ne consid\'erer que les distances
entre \'etats purs.

\subsection*{Espace des \'etats purs}

Toute $C^*$-\alg $\aa$ de dimension finie est isomorphe \`a une
somme directe finie d'alg\`ebres de matrices\cite{goodearl} \`a
entr\'ees complexes si $\aa$ est une \alg complexe, \`a entr\'ees
r\'eelles, complexes ou quaterni\-oniques si $\aa$ est une \alg
r\'eelle. On se limite aux alg\`ebres complexes de sorte que
\begin{equation} \label{algfini}
\aa = \underset{k=1}{\overset{N} \bigoplus}\, \aa_k
\end{equation}
o\`u $k, N \in\nn$, $\aa_k =\cc$ ou $M_n(\cc)$. On rappelle que
l'ensemble $\pp(\aa)$ des \'etats purs de $\aa$ jouent le r\^ole
de "points" pour l'espace noncommutatif. On montre
alors\cite{these}

{\lem  \label{etatsommedirecte} Soient $\aa_1$, $\aa_2$ deux
$C^*$-alg\`ebres, alors $\pp(\aa_1 \oplus \aa_2) = \pp(\aa_1) \cup
\pp(\aa_2)$.}
\newline

\noindent Ce lemme appliqu\'e r\'ecursivement sur $\aa$ donne
\begin{equation}
\label{etatpurespacefini} \pp(\aa) = \underset{k= 1}{\overset{N}
\bigcup} \pp(\aa_k),
\end{equation}
si bien que pour connaitre $\pp(\aa)$ il suffit de connaitre les
\'etats purs de $\cc$ et $M_n(\cc)$.

L'\'etat pur de $\cc$ n'est autre que l'identi\'e et on montre que
tout \'etat pur de $M_n(\cc)$ est associ\'e \`a un vecteur
complexe, de dimension $n$ et de norme $1$. Deux tels vecteurs
d\'efinissent le m\^eme \'etat si et seulement si ils sont \'egaux
\`a une phase pr\`es. Autrement dit
$$\pp(M_n(\cc)) = \cc P^{n-1}$$
 o\`u $\cc P^{n-1}$ d\'esigne
{\it l'espace projectif complexe} de dimension $n-1$ (espace des
vecteurs complexes de dimension $n$ \'egaux \`a une phase pr\`es).

 Les espaces finis apparaissent comme des espaces de $N$ points ($N$ d\'esigne dans (\ref{algfini}) le
nombre de composantes de $\aa$) muni chacun d'une fibre  identique
\`a $\pp(\aa_k)$. Lorsque $N=1$ on parle d'espace \`a un point. Le
cas $\aa=\cc$ est sans int\'eret. L'exemple le plus simple est
$M_n(\cc)$ repr\'esent\'e de mani\`ere irr\'eductible sur $\cc^n$.

\subsubsection*{ L'exemple de $M_2(\cc)$}

Pour $n=2$, l'espace des \'etats purs $\mathbb{C}P^{1}$ est
isomorphe \`a la sph\`ere $S^2$. Un isomorphisme explicite est
donn\'e par la projection de Hopf qui \`a tout vecteur complexe
$\xi$ de dimension deux norm\'e \`a une phase pr\`es associe le
point $p_\xi$ de $S^2$ \-- vue comme une surface dans $\rr^3$ \--
de coordonn\'ees cart\'esiennes
\begin{equation}
\label{hopf} x_{\xi}\doteq 2\re(\xi_1\bar{\xi}_2), \quad
y_{\xi}\doteq 2\im(\xi_1 \bar{\xi}_2)\,  \text{ et } \,
z_{\xi}\doteq  |\xi_1|^2 -|\xi_2|^2.
\end{equation}
On dira que deux \'etats $\xi$, $\zeta$ sont de m\^eme altitude
quand $z_{\xi} = z_{\zeta}$. On suppose que les deux valeurs
propres $D_1$, $D_2$ de l'op\'erateur de Dirac sont distinctes
(sinon $D$ est proportionnel \`a l'identit\'e et les distances
sont toutes infinies), les distances sont alors ais\'ement
calculables\cite{finite}

{\prop  La distance entre deux \'etats purs $\xi, \zeta$ est finie
si et seulement si ils sont de m\^eme altitude. Alors la distance
non commutative est la distance euclidienne sur le cercle \`a un
facteur multiplicatif pr\`es
$$d(\xi, \zeta)={\frac{2}{|D_1-D_2|}}\sqrt{(x_{\xi} - x_{\zeta})^2+(y_{\xi}-y_{\zeta})^2}.$$}

\subsection*{Espace \`a deux points}\label{deuxpoints}

Pour $N=2$ l'espace le plus simple correspond \`a l'alg\`ebre
$\mathcal{A}=M_{n}(\mathbb{C})\oplus\mathbb{C}$ represent\'ee par
une matrice diagonale par bloc sur
$\mathcal{H}=\mathbb{C}^{n}\oplus\mathbb{C}$
\begin{equation}
\label{rep2points} a=\dm{cc}x & 0\\ 0 &y\fm,
\end{equation}
avec $x \in M_n(\mathbb{C})$ et $y\in \mathbb{C}$. En prenant pour
op\'erateur de Dirac une matrice $D$ autoadjointe quelconque, les
calculs sont ardus.  En revanche, la prise en compte des axiomes
de la g\'eom\'etrie non commutative, en restreignant le choix de
l'op\'erateur de Dirac, permet de mener les calculs \`a terme.

Trois des axiomes,  relatifs \`a l'analyse fonctionnelle, sont
syst\'ematiquement v\'erifi\'es par les triplets spectraux
finis\cite{finite}. La dualit\'e de Poincar\'e est discut\'ee de
mani\`ere g\'en\'erale pour les triplets finis dans
ref.[\citelow{finite}]. Restent la r\'ealit\'e, la condition
d'ordre un et l'orientabilit\'e. Noter que pour une alg\`ebre de
dimension finie, la dimension spectrale est nulle. Il faut donc
montrer qu'il existe (r\'ealit\'e) un op\'erateur antilin\'eaire
$J= J^*= J^{-1}$ de $\hh$ dans lui-m\^eme tel que $J^2= \ii$, $[a,
JbJ^{-1}]=[J, \Gamma]=[J,D]=0.$ Tout \'el\'ement de $C_0(\aa,
\aa\ot \aa^{\circ})$ est un cycle. Les g\'en\'erateurs de
$Z_0(\aa, \aa\ot \aa^{\circ})$ sont
 les \'el\'ements de $\aa\ot \aa^{\circ}$. Il doit donc exister $a^i$, $b_i$ dans $\aa$ tels que  (orientabilit\'e) la graduation s'\'ecrive $\Gamma = a^i J b_i
J^{-1}$. Enfin l'op\'erateur de Dirac satisfait (condition du
premi\`ere ordre)  $[[D,a], J b J^{-1}] = 0$. Rappelons que par
d\'efinition la graduation commute avec $\aa$ et anticommute avec
$D$.

Si $\hh = \cc^{n+1}$, $J$ apparait comme une matrice unitaire
compos\'ee avec la conjugaison complexe, $J = U\circ c$. La
relation de commutation $[a,JbJ^{-1}] = 0$ s'\'ecrit $[a, U\bar{b}
U^*] = 0$ ce qui ne peut \^etre vrai pour tout $a$ et $b$ puisque
l'alg\`ebre n'est pas commutative. En revanche si $\aa$ est
repr\'esent\'ee sur $\hh= M_{n+1}(\cc)$ par simple multiplication
matricielle, alors un $J$ possible est l'op\'erateur  d'involution
puisque $J^2=\ii$,
$$[a, JbJ^{-1}]\psi = aJbJ^{-1}\psi - JbJ^{-1}a\psi = a J b \psi^* - Jb \psi^* a^* = a \psi b^* - a \psi b^* = 0$$
et on v\'erifie, pour n'importe quel op\'erateur de Dirac,
$$[[D,a], J b J^{-1}] \psi = [D,a] \psi b^* - [D,a]\psi b^* =0.$$
Pour que $[D,J]=0$, on peut prendre
$$D\psi = \Delta \psi + \psi \Delta$$
o\`u $\Delta=\Delta^*\in M_{n+1}(\cc)$. En effet $[D,J]\psi =
D\psi^* - JD\psi = \Delta\psi^* + \psi^*\Delta -J(\Delta\psi +
\psi\Delta) = 0.$

Concernant la graduation, le choix le plus simple est de prendre
$b_i = a^i = 0$ sauf $b_1=a^2 =\ii$ et $a^1 = b_2= K$ o\`u
$$K \doteq  \dm{cc} \ii_n & 0 \\ 0 & -1 \fm$$
n'est autre que la graduation de $\cc^{n+1}$. Ainsi $\Gamma \psi =
K\psi + \psi K$ et  $[\Gamma, J]=0$. Comme $K$ commute avec tout
$a\in\aa$, on v\'erifie que
$$[\Gamma,a]\psi = \Gamma a \psi - a\Gamma\psi = Ka\psi + a\psi K - aK\psi - a\psi K  =0.$$
Enfin,
$$(D\Gamma + \Gamma D)\psi = D(K\psi + \psi K) + \Gamma(\Delta\psi + \psi\Delta) = (\Delta K + K\Delta)\psi + \psi(\Delta K + K\Delta)$$
est nul pour tout $\psi$ si et seulement si $(\Delta K +
K\Delta)=0$. $\Delta$ s'\'ecrit donc, selon la graduation de
$\cc^{n+1}$,
$$
\Delta=\dm{cc} 0_n& m\\ m^{*}&0 \fm
$$
o\`u  $m$ un vecteur non nul de $\cc^n$.

A priori, repr\'esenter l'alg\`ebre sur $M_{n+1}(\cc)$ ne facilite
pas le calcul de la norme du commutateur $[D, a]$. Cependant la
 norme d'op\'erateur sur $M_{n+1}(\cc)$ est \'egale\cite{murphy} \`a la norme  d'op\'erateur sur $\cc^{n+1}$ si bien que, pour le calcul des distances, tout ce
passe comme si  on travaillait avec le triplet spectral $(\aa,
\hh=\cc^{n+1}, \Delta)$ au lieu de $(\aa, M_{n+1}(\cc), D).$ Les
distances sont alors facilement calculables.\cite{finite}

{\prop Si $\xi$, $\zeta$ sont tels que $\xi_j= e^{i\theta}\zeta_j$
pour tout $j\in[2,n]$,
$$d(\xi, \zeta) = \frac{2}{\norm{m}} \sqrt{1 - \abs{\scl{\xi}{\zeta}}^2}.$$
Par ailleurs $w_c$ est \`a distance infinie de tous les \'etats
purs, except\'e l'\'etat correspondant au vecteur $e_1=\dm{c} 1
\\ 0\fm$ et
$$d(\oc, \omega_{e_1}) = \frac{1}{\norm{m}}. $$}

Appliquons ces r\'esultats \`a $M_2(\cc)\oplus \cc$. L'espace des
\'etats purs est l'union disjointe de la sph\`ere $S^2$ et du
point $\oc$. Le point isol\'e, infiniment distant de tous les
autres, correspond au vecteur
$$\dm{c} 0 \\ 1\fm$$
qui, par la fibration de Hopf, est envoy\'e sur le p\^ole sud
$(0,0, -1)$ de la sph\`ere. Le point correspondant \`a $e_1$ est
le p\^ole nord, et c'est le seul point qui se trouve \`a distance
finie de $\oc$. On montre que les conditions sur la finitude des
autres distances sont identiques \`a celles du cas \`a $1$ point
et on retrouve que la distance sur des plans de m\^eme altitude
est, \`a une constante pr\`es, la distance euclidienne du cercle.

 A noter que l'ajout du point $\oc$ donne une orientation \`a la
 sph\`ere $S^2$: dans l'espace \`a un point rien ne permet de distinguer les deux
points isol\'es, tandis que dans l'espace \`a deux points le
p\^ole sud est par d\'efinition l'unique point isol\'e.

\chapter{Le mod\`ele standard}

\section{Fluctuation de la m\'etrique et transformation de jauge}

\subsection*{Transformation de jauge}

Soit $\psi$ le champ repr\'esentant un fermion de charge $e$. Le
Lagrangien libre conduisant \`a l'\'equation de Dirac
$$ (i\gamma^{\mu}\partial_\mu + m)\psi = 0$$
est
\beq \label{lagrangien}
 L = \bar{\psi} (i\gamma^{\mu}\partial_\mu  + m)\psi
\eeq o\`u $\bar{\psi}$ d\'esigne le conjugu\'e complexe de $\psi$.
Ce lagrangien est invariant sous la transformation de jauge {\it
globale}
$$\psi\rightarrow e^{-ie\alpha}\psi\quad\quad \bar{\psi}\rightarrow
e^{ie\alpha}\bar{\psi}
$$
o\`u $\alpha$ est un nombre r\'eel. Pour obtenir un lagrangien
invariant sous une transformation de jauge {\it locale}, c'est à
dire telle que le coefficient $\alpha=\alpha(x)$ depende du point
$x$ de l'espace $M$ consid\'er\'e,
$$\psi\rightarrow e^{-ie\alpha(x)}\psi\quad\quad \bar{\psi}\rightarrow
e^{ie\alpha(x)}\bar{\psi}
$$
il convient d'ajouter au lagrangien libre un terme de couplage au
potentiel vecteur $A_\mu$. On v\'erifie de la sorte que \beq
\label{lagrangieninteraction} L =
\bar{\psi}(i\gamma^{\mu}(\partial_\mu - ieA_\mu) + m)\psi \eeq
 est bien invariant de jauge locale pour peu que le
potentiel vecteur se transforme selon
$$A_\mu \rightarrow A_\mu - \partial_\mu \alpha(x).
$$
Le groupe de jauge est ici le groupe $U(1)$ car $e^{i\alpha(x)}\in
U(1)$ en tout point $x$. Cette construction se g\'en\'eralise \`a
un groupe de Lie $G$ quelconque, de g\'en\'erateurs $T_a$,  en
\'ecrivant que le langrangien (\ref{lagrangieninteraction}) est
invariant sous la transformation de jauge locale
$$\psi\rightarrow g(x)\psi\quad\quad \bar{\psi}\rightarrow
\bar{\psi}g^{-1}(x)
$$
o\`u $g(x) = e^{\theta^a(x)}T_a \in G$ et $\theta^a$ sont des
fonctions r\'eelles, d\`es lors que le potentiel de jauge
$$A_\mu(x) = A_\mu^a(x) T_a$$
o\`u  $A_\mu^a$ sont des fonctions r\'eelles, se transforme selon
$$A_\mu \rightarrow g A_\mu g^{-1} + g\partial_\mu g^{-1}.$$
G\'eom\'etriquement, le potentiel de jauge $A_\mu$ s'interpr\`ete
comme la forme locale d'une connexion sur le fibr\'e vectoriel, de
base $M$, associ\'e \`a $G$.

\subsection*{Connexion hermitienne}

Les th\'eories de jauge, du type Yang-Mills, sont construites sur
un fibr\'e vectoriel o\`u les fibres sont le support d'une
repr\'esentation du {\it groupe de jauge} de l'interaction. De la
m\^eme mani\`ere qu'\`a un espace compact $X$
 est associ\'ee l'alg\`ebre $C(X)$
de ses fonctions continues, \`a tout fibr\'e vectoriel
$E\rightarrow X$ est associ\'e le module de ses sections continues
$\Gamma(E)$ d\'efini en (\ref{modulesection}). C'est un module sur
$C(X)$ qui est fini et projectif [\citelow{jgb}, {\it Prop. 2.9}].
La d\'efinition d'un module projectif fini est donn\'ee dans la
section I.II.4  (\'enonc\'e de la condition de finitude); de
toutes ses propri\'et\'es nous retiendrons celle-ci: tout module
projectif fini sur $C(X)$  est le module des sections continues
d'un fibr\'e vectoriel sur $X$. Ce th\'eor\`eme, du \`a  Serre et
Swan, est le pendant pour les fibr\'es vectoriels du th\'eor\`eme
de Gelfand. Comme pour le couple espace compact/$C^*$-alg\`ebre
commutative, on montre que la cat\'egorie des fibr\'es vectoriels
sur un espace compact $X$ est \'equivalente \`a la cat\'egorie des
modules projectifs sur $C(X)$. Ainsi un module projectif fini sur
l'alg\`ebre $\aa$ d'un triplet spectral r\'eel $(\aa, \hh, D,
\Gamma, J)$ est un bon candidat pour jouer le r\^ole de fibr\'e
vectoriel pour la g\'eom\'etrie en question, et servir de support
\`a la formulation non commutative d'une th\'eorie de jauge.

Dans une th\'eorie de jauge, le {\it potentiel de jauge} \-- le
quadrivecteur potentiel pour l'\'electroma\-gn\'etisme par exemple
\-- est la forme locale d'une connexion, une transformation de
jauge correspondant \`a un changement de connexion. En
g\'eom\'etrie non commutative, la connexion\cite{gravity} est
d\'efinie par analogie avec la formule (\ref{connexion}). Au lieu
d'une vari\'et\'e $M$, on se donne un triplet spectral $(\aa, \hh,
D)$. $\ginf(E)$ est remplac\'e par un $\aa$-module projectif fini
$\ee$. La proposition \ref{commutprop} sugg\`ere que les
$1$-formes de la g\'eom\'etrie $(\aa, \hh, D)$ soient
g\'en\'er\'ees par des \'el\'ements du type $[D,a]$. L'ensemble
$\Omega^1(M)$ des sections de $T^*M$ est un $C(M)$-module. On
demande donc que l'ensemble $\Omega^1_D$ des $1$-formes de la
g\'eom\'etrie $(\aa, \hh, D)$ soit un $\aa$-module. Autrement dit
\begin{equation}
\label{omega1d} \Omega^1_D \doteq \left\{ a^i [D, b_i]\,, \; a^i,
b_i \in \aa\right\}.
\end{equation}
{\defi Soit $(\aa, \hh, D)$ un triplet spectral. Une {\it
connexion} sur un $\aa$-module projectif fini $\ee$ est une
application $\aa$-lin\'eaire $\del:\, \ee \mapsto \ee\ot_{\aa}
\Omega^1_D$ satisfaisant la r\`egle de Leibniz
$$\del (sa) = (\del s) a + s\ot [D,a]$$
pour tout $a\in\aa, s\in \ee$.}
\newline

Lorsque qu'un fibr\'e vectoriel $E\rightarrow X$ est muni d'un
produit scalaire fibre \`a fibre, le module $\Gamma(E)$ h\'erite
d'une structure hermitienne \`a valeur dans $C(X)$:
$$(\sigma_1 \lvert \sigma_2) (x) = \scl{\sigma_1(x)}{\sigma_2(x)}.$$ Adapt\'ee \`a un module (par convention \`a droite)
sur une $C^*$-alg\`ebre $\aa$ quelconque, la structure hermitienne
d\'efini un {\it $C^*$-module.}

{\defi Un $C^*$-module sur une $C^*$-alg\`ebre $\aa$ est un espace
vectoriel $\ee$ qui est aussi un $\aa$-module (pas forc\'ement
projectif fini) muni d'un couplage $\ee\times \ee \rightarrow \aa$
tel que
\begin{eqnarray*}
(r \vert s+t) &=& (r \vert s) + (r \vert t),\\
(r \vert s a) &=& (r\vert s) a,\\
(r\vert s)    &=& (s \vert r)^*,\\
(s\vert s)    &\rangle& 0 \text{ pour } s\neq 0
\end{eqnarray*}
o\`u $r,s,t \in \ee$ et $a\in\aa$, tel que $\ee$ soit complet pour
la norme
$$\norm{s}\doteq\sqrt{\norm{(s\vert s)}}.$$}

\noindent Les modules plein de la d\'efinition \ref{morita} de
l'\'equivalence de Morita sont des $C^*$-modules.

Quand un $\aa$-module projectif fini $\ee$ est aussi un
$C^*$-module \-- $\aa$ est une $C^*$-alg\`ebre \-- se pose la
question de la compatibilit\' e de la connexion avec la structure
hermitienne. L'\'equivalent non commutatif de la connexion de
Levi-Civita est une {\it connection hermitienne}, ie. une
connexion satisfaisant la version non commutative de (\ref{lvg}),
\`a savoir
\begin{equation}
\label{connexhermit} ( s \lvert \del r) - (\del s \lvert r) = [D,
(s\lvert r)].
\end{equation}
Pr\'ecisons que si $\del s = s^i \ot \varpi_i$, $s^i\in \ee$,
$\varpi^i\in\Omega^1_D$, alors
$$(\del s \lvert r) \doteq {\varpi_i}^* (s^i\lvert r )\; \text{ et }\; ( r \lvert \del s) \doteq (r \lvert s^i)\varpi_i.$$
La diff\'erence d'un signe $-$ entre (\ref{lvg}) et
(\ref{connexhermit}) provient de la d\'efinition $d a \doteq
[D,a]$, puisqu'alors $d(a^*) = - (da)^*$. Un th\'eor\`eme
fondamental de la g\'eom\'etrie riemannienne indique que pour
toute vari\'et\'e (pseudo)-riemannienne, il existe une unique
connexion compatible avec la m\'etrique et de torsion nulle. Pour
les $C^*$-modules projectifs fini, on un th\'eor\`eme du m\^eme
ordre, qui repose sur le fait que tout module projectif fini sur
$\aa$ est de la forme
\begin{equation}
\label{empf} \ee = e\aa^N
\end{equation}
o\`u $\aa^N$ d\'esigne le $\aa$-module des vecteurs colonnes de
dimension $N$ \`a entr\'ee dans $\aa$, et $e = e^2\in M_N(\aa)$.
Tout \'el\'ement $s$ d'un $\aa$-module projectif fini est un
$\aa$-vecteur colonne et, puisque $\Omega^1_D$ est un
$\aa$-module, $\del s\in \ee\ot_\aa\Omega^1_D$ est un vecteur \`a
entr\'ee dans $\Omega^1_D$. On note $\xi\in\aa^N$ le vecteur de
composante $\xi_j\in\aa$ tel que $s=e\xi$, et $d\xi$ le vecteur de
composante $[D, \xi_i]\in \Omega^1_D$. On montre alors que
l'ensemble des connexions hermitiennes est un espace affine.

{\prop\label{connexhermiti} Soit $\ee\simeq e\aa^N$ un
$C^*$-module projectif fini. La structure hermitienne de $\ee$ est
induite par la structure hermitienne canonique
 de $\aa^N$. Sur ce module, toutes les connexions hermitiennes sont donn\'ees par
$$\del (e\xi) = d(e\xi) + eAe\xi$$
o\`u $A\in M_N(\Omega^1_D)$ est une matrice hermitienne.}
\newline

Toute endomorphisme inversible $\alpha$ de $\ee$ d\'efinit un
endomorphisme de l'espace des connexions
\begin{equation}
\label{connexend} \del \mapsto (\alpha\ot \ii) \del \alpha^{-1}.
\end{equation}
On peut choisir de faire agir un endomorphisme de $\ee$ sur
l'espace des connexions autrement, mais l'action (\ref{connexend})
permet de caract\'eriser facilement un certain type
d'endomorphisme qui pr\'eserve l'hermicit\'e. Un endomorphisme
$\aa$-lin\'eaire $\alpha$ de $\ee$ poss\`ede un adjoint s'il
existe un endomorphisme $\alpha^*$ tel que
$$(r \lvert \alpha s ) = (\alpha^* r \lvert s)$$
pour tout $r,s \in\ee$. On note $\text{End}_A
(E)$\index{endea@$\text{End}_A (E)$} l'alg\`ebre des
endomorphismes
 avec adjoint (c'est une $C^*$-alg\`ebre pour la norme d'op\'erateur [\citelow{jgb},{\it Th. 3.1}]. Un tel endomorphisme est {\it
unitaire} s'il pr\'eserve la structure hermitienne
$$(\alpha r \lvert \alpha s) = (\ r \lvert s),$$
c'est \`a dire si $\alpha^*\alpha = \alpha\alpha^* = \ii_{\ee}$
(l'endomorphisme identit\'e). Le groupe des endomorphismes
unitaire est not\'e ${\cal U}(\ee)$.\index{ue@${\cal U}(\ee)$} On
montre alors\cite{connes} que si $\del$ est une connexion
hermitienne sur $\ee$ et $u\in{\cal U}(\ee)$, alors $(u\ot\ii)\del
u^*$ est une connexion hermitienne. D'o\`u la d\'efinition d'une
{\it transformation de jauge}.

{\defi L'action de ${\cal U}(\ee)$ sur les connexions hermitiennes
est appel\'ee transformation de jauge.}
\newline

\noindent La matrice $A$ de la proposition \ref{connexhermiti} est
l'\'equivalent non commutatif du potentiel de jauge.

\subsection*{Op\'erateur de Dirac covariant}

Etant donn\'es une g\'eom\'etrie $(\aa, \hh, D, J, \Gamma)$ et un
$\aa$-module projectif fini $\ee$, on peut construire des
connexions sur $\ee$. L'interpr\'etation g\'eom\'etrique de ces
connexions, c'est \`a dire leur influence sur la g\'eom\'etrie
$(\aa, \hh, D)$, passe par la construction d'un nouveau triplet
spectral.

Tout \'el\'ement $s$ d'un $\aa$-module projectif fini est un
$\aa$-vecteur colonne. On note $\bar{s}$ le $\aa$-vecteur ligne
correspondant. L'ensemble des $\bar{s}$ pour $s\in \ee$ est un
$\aa$-module projectif \`a gauche, not\'e $\bar{\ee}$, o\`u
l'action de $\aa$ est
$$a\bar{s} \doteq \overline{sa^*}.$$

{\prop Soit $(\aa, \hh, D, \Gamma)$ un triplet spectral r\'eel de
dimension $n$ et $\del$ une connexion hermitienne sur un
$\aa$-module projectif finie $\ee$. Soit
\begin{eqnarray*}
\tilde{\aa}&\doteq& \text{ End}_A(\ee),\\
\tilde{\hh}&\doteq& \ee \ot_\aa \hh \ot_\aa \bar{\ee}
\end{eqnarray*}
et l'op\'erateur $\tilde{D}$ agissant sur $\tilde{\hh}$ par
$$\tilde{D} (s\ot \psi\ot \bar{r}) \doteq (\del s)\psi\ot\bar{r} + s\ot D\psi\ot\bar{r} + s\ot \psi\overline{\del r}.$$
Alors $(\tilde{\aa}, \tilde{\hh}, \tilde{D},  \tilde{J},
\tilde{\Gamma})$ avec
\begin{eqnarray*}
\tilde{J}(s\ot\psi\ot\bar{r}) &\doteq& r\ot J\psi \ot \bar{s},\\
\tilde{\Gamma}(s\ot\psi\ot\bar{r}) &\doteq& s\ot \Gamma\psi \ot
\bar{r}
\end{eqnarray*}
est un triplet spectral r\'eel de dimension $n$.}
\newline

\noindent L'action de $\del s = s^i\ot\varpi_i$ sur $\hh$ est
d\'efini en voyant $\varpi_i$ comme un op\'erateur sur $\hh$ via
la d\'efinition (\ref{omega1d}) de $\Omega^1_D$
$$(\del s)\psi = s^i \ot \varpi_i\psi.$$
De m\^eme on d\'efinit $\psi\overline{\del s} = \psi
\overline{s^i\ot\varpi_i} \doteq  J\varpi_i J^{-1}
\psi\ot\bar{s^i}.$

Quand $\tilde{\aa}\neq\aa$ les deux g\'eom\'etries sont
difficilement comparables puisqu'elles ne reposent pas sur le
m\^eme espace des \'etats. En revanche, si on choisit le
$\aa$-module trivial $\ee=\bar{\aa}=\aa$,  on obtient
$\tilde{\aa}= \aa$, $\tilde{\hh}=\hh$ et $\tilde{D}= D + A +
JAJ^{-1}.$
{\defi L'op\'erateur $D_A\doteq D + A + JAJ^{-1}$ est appel\'e
op\'erateur de Dirac covariant.}
\newline

\noindent L'emploi du terme covariant se justifie en remarquant
que l'action d'un unitaire $u\in{\cal U}(\aa)$, par la
modification de la connexion, induit une transformation de $D_A$
en
$$
D_{A'} = D + A' + JA'J^{-1},
$$
o\`u $A'\doteq u Au^* + u[D,  u^*] $. Autrement dit sous une
transformation de jauge, $A$ se transforme selon
$$A\mapsto   u Au^*+ u[D, u^*].$$
qui est bien la  loi de transformation du potentiel vecteur en
\'electromagn\'etisme
$$A\mapsto uAu^{-1} + udu^{-1}.$$

Comme a priori $[D_A, a]\neq [D,a]$ pour un $a$ quelconque de
$\aa$, le remplacement de $D$ par $D_A$, c'est \`a dire le passage
d'une th\'eorie \`a connexion nulle \`a une th\'eorie covariante,
induit une perturbation de la m\'etrique appel\'ee {\it
fluctuation interne de la m\'etrique}. En particulier, si la
courbure associ\'ee \`a la d\'efinition non commutative de la
connexion est non nulle, alors on peut de mani\`ere imag\'ee voir
les fluctuations internes de la m\'etrique comme t\'emoignage de
la "courbure de la non commutatitivit\'e", sans \'equivalent
commutatif puisqu'en ce cas $A$ est nul.

\section{Produit de g\'eom\'etries}

\subsection*{Produit de triplets spectraux}

Le produit tensoriel d'un triplet spectral r\'eel pair
$T_I=(\aa_I, \hh_I, D_I,\pi_I)$ muni d'une chiralit\'e $\Gamma_I$,
par le triplet spectral r\'eel $T_E=(\aa_E, \hh_E, D_E,\pi_E)$ est
le triplet spectral $T_I\ot T_E\doteq(\aa',\hh',D')$ d\'efini par
\begin{equation*}
\label{pdt0} \aa'\doteq \aa_I\otimes\aa_E,\quad
\hh'\doteq\hh_I\otimes\hh_E,\quad  D'\doteq D_I\otimes\ii_E +
\Gamma_I\otimes D_E.
\end{equation*}
La repr\'esentation est $\pi'\doteq \pi_I\ot \pi_E$ (dans ce
chapitre nous n'utiliserons ni la chiralit\'e ni la structure
r\'eelle du triplet produit mais toutes deux sont d\'efinies, cf
[\citelow{vanhecke}]). Dans la mesure o\`u les triplets spectraux
ne forment pas un espace vectoriel, la notation $T_I\ot T_E$ est
essentiellement une convention. Ce produit est commutatif car
lorsque $T_E$ est pair et muni d'une chiralit\'e $\Gamma_E$,
alors le triplet spectral $T_E\otimes T_I\doteq(\aa,\hh,D)$ est
\'egalement d\'efini (il suffit de permuter les facteurs)
\begin{equation}
\label{pdt1} \aa\doteq \aa_E\otimes\aa_I,\quad
\hh\doteq\hh_E\otimes\hh_I,\quad  D\doteq D_E\otimes\ii_I +
\Gamma_E\otimes D_I\, ,
\end{equation}
$\pi=\pi_E\ot\pi_I$ et il est \'equivalent \`a $T_I\otimes T_E$
via l'op\'erateur unitaire
\begin{equation*}
U\doteq \frac{\ii_I+ \Gamma_I}{2}\otimes\ii_E + \frac{\ii_I-
\Gamma_I}{2}\otimes\Gamma_E\,.
\end{equation*}

En physique ce produit tensoriel est utilis\'e pour d\'ecrire un
espace continu dont chaque point est muni d'une fibre discr\`ete.
Dans le mod\`ele standard l'espace interne $T_I$ est choisie de
mani\`ere \`a ce que le groupe des unitaires de $\aa_I$, modulo le
rel\`evement aux spineurs\cite{spingroup,farewell}, soit le groupe
de jauge des interactions. $\aa_I$ est une \alg de matrices,
$\hh_I$ est l'espace des fermions et l'op\'erateur de Dirac
interne a pour coefficients les masses des fermions,
\'eventuellement pond\'er\'ees par la matrice unitaire de
Cabibbo-Kobayashi-Maskawa.

\subsection*{1-forme dans un produit de g\'eom\'etries}

Dans un produit de g\'eom\'etries, Les 1-formes sont donn\'ees
par\cite{kt,schucker}
$$
\Omega^1=\Omega^1_E\ot \Omega^0_I +
\chi_E\Omega^0_E\ot\Omega^1_I\,,
$$
o\`u $\Omega^0_E=\aa_E$ est l'ensemble des 0-formes de $\aa_E$,
les autres termes \'etant d\'efinis de mani\`ere analogue.
 Quand $T_E$ est le triplet spectral d'une vari\'et\'e,
$$
\Omega_E^1\ni f^j[-i\ds,g_j\ii_E]= -if^j(\gamma^m \partial_\mu
g_j)=-i\gamma^m f_\mu\,,
$$
o\'u $f^j,g_j,f_\mu\doteq f^j\partial_\mu g_j \in\cinf$. Une
1-forme du triplet total est
$$
\Omega^1\ni -i\gamma^m f_\mu^i \ot a_i - \gamma^5 h^j\ot m_j$$
o\`u $a_i\in\aa_I$, $h^j\in\cinf$, $m_j\in\Omega_I^1$. Un
potentiel vecteur est donn\'e par
\begin{equation}
\label{h} A = -i\gamma^m \ot A_\mu - \gamma^5\ot H
\end{equation}
avec $A_\mu\doteq {f^i}_\mu a_i$ un champ de vecteur (sur $\mm$)
\`a valeur dans les \'el\'ements anti-adjoints de $\aa_I$ et
$H\doteq h^j m_j$ un champs scalaire \`a valeur dans $\Omega^1_I$.
Pour une alg\`ebre de matrices (ou une somme directe d'alg\`ebres
de matrices), les \'el\'ements anti-adjoints forment l'alg\`ebre
de Lie du groupe des unitaires. Ce groupe de Lie repr\'esente le
groupe de jauge de la th\'eorie, donc  $A_\mu$ est un potentiel de
jauge. Dans [\citelow{gravity}] une formule est donn\'ee  pour les
fluctuations de la m\'etrique dues \`a  $A_\mu$. Ici nous nous
int\'eressons aux fluctuations provenant uniquement du champ
scalaire $H$, et {\bf on suppose que
 ${\mathbf A_\mu=0}$.} On calcule alors que
\begin{equation}
\label{daa} [D_A,a]= [D - \gamma^5\ot H,a].
\end{equation}

Dor\'enavant on \'ecrit  $D_A\doteq D - \gamma^5\ot H$. Pour ne
pas alourdir les notations, on d\'esigne toujours par $d$ la
distance associat\'ee au triplet $(\aa, \hh, D_A)$. Selon
(\ref{pdt1}), une fluctuation scalaire substitue
$$D_H\doteq D_I+ H$$
\`a $D_I$. La diff\'erence essentielle est que maintenant
l'op\'erateur de Dirac $D_H$ d\'epend de $x$, de sorte que tout
point de $\mm$ d\'efinit un triplet spectral interne
$$
T_I^x\doteq (\aa_I, \hh_I, D_H(x))\,.
$$

\section{ Le mod\`ele standard.}

Le triplet spectral du mod\`ele standard (cf.
[\citelow{connes,gravity,spectral}] et [\citelow{bridge}] pour le
calcul d\'etaill\'e de la masse du boson de Higgs) est le produit
du triplet spectral r\'eel (\ref{td}), not\'e ici $T_E$,  par une
g\'eom\'etrie interne o\`u l'alg\`ebre
$$\aa_I=\hhh \oplus \cc \oplus M_3(\cc)$$
($\hhh$ d\'esigne l'alg\`ebre des quaternions) est represent\'ee
sur
$$\hh_I=\cc^{90}=\hh^P \oplus \hh^A= \hh_L^P \oplus \hh_R^P \oplus \hh_L^A
\oplus \hh_R^A\,.$$ La base de $\hh_L^P=\cc^{24}$ est donn\'ee par
les fermions gauches
$$
\dm{c} u\\d \fm_L,\;\dm{c} c\\s \fm_L,\;\dm{c} t\\b \fm_L,\;\dm{c}
\nu_e\\e \fm_L,\;\dm{c} \nu_\mu\\\mu \fm_L,\;\dm{c} \mu_\tau\\
\tau \fm_L,
$$
 et la base de $\hh_R^P=\cc^{21}$ est form\'ee des fermions droits
$u_R,\, d_r,\, c_R,\, s_R,\, t_R,\, b_R \text{ et }\, e_R,\,
\mu_R,\, \tau_R$ (le mod\`ele a \'et\'e construit du temps o\`u
les neutrinos n'avaient pas de masse). L'indice de couleur des
quarks est omis.
 $\hh_R^A$ et $\hh_L^A$ correspondent  aux antiparticules. $(a\in\hhh,\; b\in\cc,
c\in M_3(\cc))$ est repr\'esent\'e par
\begin{equation*}
\label{repms} \pi_I(a,b,c)\doteq \pi^P (a,b) \oplus \pi^A (b,c)
\doteq \pi_L^P(a) \oplus \pi_R^P(b) \oplus \pi_L^A(b,c) \oplus
\pi_R^A (b,c)
\end{equation*}
o\`u, en \'ecrivant $B\doteq\dm{cc} b~&~0\\0~&~\bar{b}\fm\in\hhh$
et $N$ le nombre de g\'en\'erations de fermions,
\begin{eqnarray*}
\pi_L^P(a)\doteq a\ot\ii_N\ot\ii_3\,  \oplus\, a\ot\ii_N\,,\qquad&
&
\pi_R^P(b) \doteq B \ot \ii_N \ot \ii_3\, \oplus \, \bar{b}\ot \ii_N\,,\\
\pi_L^A(b,c)\doteq \ii_2\ot\ii_N\ot c \,\oplus\,
\bar{b}\ii_2\ot\ii_N\,,\qquad & &
\pi_R^A(b,c)\doteq\ii_2\ot\ii_N\ot c \, \oplus \, \bar{b}\ii_n\,.
\end{eqnarray*}
On d\'efinit une structure r\'eelle
$$
J_I= \dm{cc} 0~ & ~\ii_{15N}\\ \ii_{15N}~ & ~0\fm \circ\, C
$$
et un op\'erateur de Dirac interne
$$
D_I\doteq\dm{cc} D_P & 0\\ 0& \;\bar{D_P} \fm= \dm{cc} D_P~ & 0 \\
0&0\fm + J_I \dm{cc} D_P~ & 0 \\ 0&0\fm J^{-1}_I$$ dont les
entr\'ees sont les matrices $15N\times 15N$
$$D_P \doteq \dm{cc} 0&M\\ M^{*}& 0\fm,
$$
o\`u $M$ est la matrice $8N \times 7N$
\def\masseproj{\lp e_{11} \ot M_u + e_{22} \ot M_d \rp}
\def\masse{ \dm{cc}\masseproj \ot \ii_3& 0 \\0&e_2\ot M_e\fm}
\begin{equation}
\label{m} M\doteq\masse.
\end{equation}
Ici, $\{e_{ij}\}$ et $\{e_i\}$ d\'esignent les bases canoniques de
$\m2$ et $\cc^2$ respectivement. $M_u$, $M_d$, $M_e$ sont les
matrices de masse
$$M_u=\dm{ccc} m_u & 0&0 \\ 0&m_c&0\\ 0&0&m_t\fm,\quad
M_d=C_{KM}\dm{ccc} m_d & 0&0 \\ 0&m_s&0\\ 0&0&m_b\fm ,\quad
M_e=\dm{ccc} m_e & 0&0 \\ 0&m_\mu&0\\ 0&0&m_\tau\fm
$$
 dont les coefficients sont les masses des fermions \'el\'ementaires,
 \'eventuellement pond\'er\'ees par la matrice unitaire de Cabibbo-Kobayashi-Maskawa. La chiralit\'e,
 dernier \'el\'ement du triplet spectral r\'eel, est
$$
\Gamma_I= (-\ii_{8N}) \oplus \ii_{7N} \oplus (-\ii_{8N}) \oplus
\ii_{7N}\,.
$$

La g\'eom\'etrie non commutative donne une interpr\'etation du
champ de Higgs comme $1$-forme de la g\'eom\'etrie interne. Par
fluctuation scalaire,  les $1$-formes sont \'etroitement li\'ees
\`a la m\'etrique et le champ de Higgs s'interpr\`ete en effet
comme coefficient d'une m\'etrique.

Le calcul suivant est men\'e en jauge nulle $A_\mu = 0$.\cite{kk}
{\prop \label{distancems} La partie finie de la g\'eom\'etrie du
mod\`ele standard avec fluctuation interne scalaire de la
m\'etrique en jauge nulle est un  mod\`ele \`a deux couches
index\'ees par les \'etats de $\cc$ et $\hhh$. Chacune des couches
est une copie de la vari\'et\'e  riemannienne \`a spin initiale
$\mm$, munie de sa m\'etrique. La distance entre les couches est
identique \`a la distance g\'eod\'esique dans la vari\'et\'e $\mm
\times [0,1]$ de dimension 4+1, o\`u les deux copies de $\mm$
correspondent aux valeurs $0$ et $1$ de la dimension
suppl\'ementaire. La composante suppl\'ementaire de la m\'etrique,
correspondant \`a cette dimension suppl\'ementaire,  est
$$
g^{tt}(x)= \lp\abs{1+h_1(x)}^2+\abs{h_2(x)}^2\rp m_t^2
$$
o\`u $\dm{c} h_1\\h_2\fm$ est le doublet de Higgs et $m_t$ la
masse du quark top.}
\newline

A noter que quoique la distance soit identique \`a celle d'une
vari\'et\'e de dimension $4+1$, il s'agit uniquement d'une
analogie. Il n'y a pas de "points" entre les deux copies de $\mm$.
C'est pr\'ecis\'ement un grand int\'eret de la g\'eom\'etrie non
commutative: d\'ecrire un espace form\'e de deux composantes
disconnexes et pourtant \`a distance finie l'une de l'autre.
L'espace interne se comporte comme une dimension suppl\'ementaire
du point de vue de la m\'etrique, mais topologiquement il s'agit
d'une dimension suppl\'ementaire discr\`ete.

\chapter {Neutrinos massifs}

Le mod\`ele pr\'ec\'edent a \'et\'e construit pour des neutrinos
de masse nulle. Dans ce chapitre nous allons \'etudier une
mani\`ere simple d'introduire des neutrinos massifs. On obtient
ainsi une contrainte sur le nombre et la nature des neutrinos
massifs. Noter que cette mani\`ere est la modification  la plus
simple \`a apporter au mod\`ele pour introduire des neutrinos
massifs,mais elle n'est pas la seule possibilit\'e. En particulier
les contraintes qui seront mises en \'evidence, si elles se
r\'ev\'elaient fausses exp\'erimentalement, ne signifierait pas
que tout le mod\`ele est faux, mais plus simplement qu'il s'agit
de le modifier de mani\`ere moins \'el\'ementaires que celles que
nous proposons maintenant. Une approche plus syst\'ematique de la
question des neutrinos massifs en g\'eom\'etrie non commutative
peut \^etre trouv\'ee dans
[\citelow{joseneutrinos},\citelow{schelp}]

\section{Modification du triplet spectral}

 Donner une masse aux neutrinos signifie que ces derniers,
qui n'existaient qu'avec la chiralit\'e gauche, existe aussi avec
la chiralit\'e droites. Le nombres de particules \'el\'ementaires
augmentent, ce qui, dans notre mod\`ele, signifie que la dimension
de l'espace de Hilbert sur lequel est repr\'esent\'e l'alg\`ebre
interne augmente elle aussi. Supposons donc qu'on rajoute
$\alpha\leq 3$ neutrinos droits. L'espace de repr\'esentation des
particules droites devient
$$\hh^P_R \rightarrow \hh^P_R + \hh^\alpha$$
et \`a la repr\'esentation des particules droites il convient
d'ajouter la repr\'esentation des nouveaux neutrinos
$$ \Pi^P_R(a,b,c) \rightarrow \Pi^P_R(a,b,c) \oplus \Pi^{\alpha}(a,b,c).
$$
Des modifications similaires sont apport\'es \`a la partie
antiparticule de la repr\'esentation, et il convient enfin
d'ajouter \`a l'op\'erateur de Dirac internes des coefficients
correspondants aux $2\times \alpha$ dimensions suppl\'ementaires.
Enfin il faut v\'erifier que ces modifications soient compatibles
avec les axiomes de la g\'eom\'etrie non commutative.

Avant d'examiner les contraintes apport\'es par les axiomes sur
les modifications possibles pour incorporer les neutrinos massifs,
remarquons qu'en faisant l'hypoth\`ese de neutrinos droits
st\'eriles, alors leur repr\'esentation doit \^etre invariantes
par transformation de jauge. Autrment dit pour tout unitaire
$U\in\aa_I$, on demande que
$$U\Pi^{\alpha} (a,b,c)U^* = \Pi^{\alpha}(a,b,c)$$
quels que soit $a\in\hhh, b\in\cc$ et $c\in M_3(\cc)$. Ceci impose
que $\Pi^{\alpha}(a,b,c)$ soit une matrice diagonale, c'est \`a
dire que seule l'alg\`ebre $\cc$ soit repr\'esent\'ee
$$\Pi^{\alpha}(a,b,c)= \Pi^{\alpha}(b) = \text{diag} (b,
\bar{b}).
$$
\section{Dualit\'e de Poincar\'e}

 Pour des alg\`ebres de matrices,  la formule (\ref{intersection}) prend une forme simple\cite{krajew}
La matrice d'intersection pour le triplet spectral du mod\`ele
standard sans neutrinos massifs s'\'ecrit
\begin{equation}
\cap([p_i], [p_j]) = \text{Tr} \lp \Gamma_I \pi_I(p_i) J
\pi_I(p_j)\rp J^{-1})\doteq \cap_{ij}
\end{equation}
o\`u les $p_i\in \aa_I$, $i=\cc, \hhh, M_3(\cc)$, sont donn\'es
par
$$p_{\hhh} = \dm{cc} 1 & 0 \\ 0 & 1 \fm \, ,\; p_{M_3} = \dm{ccc} 1 & 0 & 0 \\ 0 & 0 & 0\\ 0 & 0 & 0
\fm \, ,\; p_{\cc} = 1,
$$
de sorte que
\begin{equation}
\label{matriceintersec} \cap = 6 \dm{ccc} 1 & -1 & 1 \\ -1 & 0 &
-1\\ 1 & -1 & 0 \fm
\end{equation}
dont le d\'eterminant est non nul. Ainsi la dualit\'e de
Poincar\'e est bien satisfaite pour la g\'eom\'etrie du mod\`ele
standard. La modification de la repr\'esentation pour des
neutrinos st\'eriles n'affecte que l'alg\`ebre $\cc$ et la matrice
d'intersection devient
\begin{equation}
\label{matriceintersecprime} \cap = 6 \dm{ccc} 6 + \underset{i=1}{\overset{\alpha}{\Sigma}} \epsilon_i & -6 & 6 \\
-6 & 0 & -6\\ 6 & -6 & 0 \fm
\end{equation}
o\`u $\epsilon_i = 2$ pour un neutrino distinct de son
antiparticule, $\epsilon_i =1$ pour un neutrino identique \`a sa
propre antiparticule. Le d\'eterminant de la matrice
d'intersection
$$ \text{ det } \cap = 36 ( 6 - \underset{i=1}{\overset{\alpha}{\Sigma}}
\epsilon_i)
$$
est nul si et seulement si $\alpha = 3$ et $\epsilon_1 =
\epsilon_2 = \epsilon_3 = 2$. Autrement dit, pour que la dualit\'e
de Poincar\'e soit satisfaite, il est n\'ecessaire qu'au moins un
neutrino soit de masse nulle, ou bien que l'un d'entre eux au
moins soit  sa propre antiparticule.

Cependant, les masses des neutrinos doivent \^etre incorpor\'ees
\`a l'op\'erateur de Dirac. Dans la base de $\cc^{90 +
\underset{i=1}{\overset{\alpha}{\Sigma}} \epsilon_i}$ labell\'ee
par les particules et les antiparticules, la masse apparait \`a la
$\text{particule}^{\text{i\`eme}}$ ligne,
$\text{antiparticule}^{\text{i\`eme}}$ colonne. Pour un neutrino
identique \`a son antiparticule, un coefficient apparait donc sur
la diagonale de l'op\'erateur de Dirac. Mais alors il est
impossible de modifier la chiralit\'e de sorte qu'elle commute
avec $ \Pi^{\alpha}(b) = \text{diag} (b, \bar{b})$ et anticommute
avec l'op\'erateur de Dirac. Autrement on ne peut incorporer que
des neutrinons distincts de leur antiparticules, et l'un d'entre
eux doit \^etre de masse nulle.

Dans ce cas, la dualit\'e de Poincar\'e est bien satisfaite, et on
montre que les autres \'el\'ements du triplet spectral peuvent
\^etre adapt\'es de mani\`ere \`a satisfaire tous les axiomes.

\end{document}